\pdfoutput=1
\documentclass[sigconf,nonacm=true]{acmart}

\AtBeginDocument{%
  }

\setcopyright{none}
\usepackage{mathpartir}
\mprset{sep=1.1em, andskip=0.4em}
\usepackage{tabularx}
\usepackage{enumitem}
\setlist{nosep,leftmargin=1.3em}
\usepackage{eso-pic}
\usepackage{tikz}
\usetikzlibrary{positioning,arrows.meta,calc}

\newif\ifdraftmark
\draftmarkfalse
\ifdraftmark
\AddToShipoutPictureFG{%
  \AtPageLowerLeft{%
    \put(\LenToUnit{0.5\paperwidth},\LenToUnit{1.05cm}){%
      \makebox(0,0){\footnotesize\itshape\color{gray}%
        Draft manuscript --- please do not distribute.}}}}
\fi
\AtBeginDocument{%
  \setlength{\abovedisplayskip}{4pt plus 1.5pt minus 1.5pt}%
  \setlength{\belowdisplayskip}{4pt plus 1.5pt minus 1.5pt}%
  \setlength{\abovedisplayshortskip}{2pt plus 1pt}%
  \setlength{\belowdisplayshortskip}{2pt plus 1pt}%
  \setlength{\textfloatsep}{8pt plus 2pt minus 2pt}%
  \setlength{\floatsep}{6pt plus 2pt minus 2pt}%
}

\newtheoremstyle{licsplain}%
  {3pt plus 1pt minus 1pt}%
  {3pt plus 1pt minus 1pt}%
  {\itshape}{\parindent}{\scshape}{.}{.5em}%
  {\thmname{#1}\thmnumber{ #2}\thmnote{ (#3)}}
\newtheoremstyle{licsdef}%
  {3pt plus 1pt minus 1pt}%
  {3pt plus 1pt minus 1pt}%
  {\normalfont}{\parindent}{\itshape}{.}{.5em}%
  {\thmname{#1}\thmnumber{ #2}\thmnote{ (#3)}}

\theoremstyle{licsplain}
\newtheorem{theorem}{Theorem}[section]
\newtheorem{lemma}[theorem]{Lemma}
\newtheorem{proposition}[theorem]{Proposition}

\newtheorem{conjecture}[theorem]{Conjecture}

\theoremstyle{licsdef}
\newtheorem{definition}[theorem]{Definition}

\newtheorem{remark}[theorem]{Remark}

\makeatletter
\renewenvironment{proof}[1][\proofname]{\par
  \pushQED{\qed}%
  \normalfont \topsep2.5\p@\@plus1.5\p@\relax
  \trivlist
  \item[\@proofindent\hskip\labelsep
        {\@proofnamefont #1\@addpunct{.}}]\ignorespaces
}{%
  \popQED\endtrivlist\@endpefalse
}
\makeatother

\DeclareRobustCommand{\code}[1]{\texttt{\small #1}}
\newcommand{\ctx}{\Gamma}
\newcommand{\tttEnv}{\mathcal{E}}
\newcommand{\tttwf}{\;\mathsf{wf}}
\newcommand{\tttctx}{\;\mathsf{ctx}}
\newcommand{\Srt}{\mathsf{Sort}}
\newcommand{\Prop}{\mathsf{Prop}}
\newcommand{\Setu}{\mathsf{Set}}
\newcommand{\Typeu}{\mathsf{Type}}
\newcommand{\lzero}{0}
\newcommand{\lsucc}{\mathsf{succ}}
\newcommand{\lmax}{\mathsf{max}}
\newcommand{\limax}{\mathsf{imax}}
\newcommand{\tttsim}{\equiv_{\mathrm{lvl}}}
\newcommand{\cumle}{\preceq_{\mathrm{lvl}}}
\newcommand{\defeq}{\equiv}
\newcommand{\subty}{\preceq}
\newcommand{\tttstep}{\triangleright}
\newcommand{\infr}{\Rightarrow}
\newcommand{\chk}{\Leftarrow}
\newcommand{\chkred}{\searrow}
\newcommand{\tttlam}[2]{\lambda #1{:}#2.\,}
\newcommand{\tttpair}[2]{\langle #1, #2 \rangle}
\newcommand{\fst}{\mathsf{fst}}
\newcommand{\snd}{\mathsf{snd}}
\newcommand{\Id}{\mathsf{Id}}
\newcommand{\refl}{\mathsf{refl}}
\newcommand{\Jelim}{\mathsf{J}}
\newcommand{\irrel}{\mathsf{irrel}}
\newcommand{\tttann}[2]{(#1 : #2)}
\newcommand{\tttinst}[2]{#1.\{#2\}}
\newcommand{\rec}{\mathsf{rec}}
\newcommand{\Nat}{\mathsf{Nat}}

\newcommand{\tttBool}{\mathsf{Bool}}
\newcommand{\tttFalse}{\mathsf{False}}
\newcommand{\tttOr}{\mathsf{Or}}
\newcommand{\tttExists}{\mathsf{Exists}}
\newcommand{\tttAcc}{\mathsf{Acc}}
\newcommand{\tttrule}[1]{\textsc{#1}}

\newcommand{\JTerm}{\mathsf{JTerm}}
\newcommand{\jlCode}{\mathsf{Code}}
\newcommand{\jlLeaf}{\mathsf{Leaf}}
\newcommand{\jlName}{\mathsf{Name}}
\newcommand{\jdecl}{\mathsf{decl}}
\newcommand{\jobs}{\mathsf{obs}}
\newcommand{\jver}{\mathsf{ver}}
\newcommand{\japp}{\mathsf{app}}
\newcommand{\jsum}{\mathsf{sum}}
\newcommand{\jinst}{\mathsf{inst}}
\newcommand{\jpair}{\mathsf{pair}}
\newcommand{\jlchk}{\mathord{!}}
\newcommand{\JB}{\mathsf{JB}}
\newcommand{\Evid}{\mathcal{E}}
\newcommand{\supp}{\mathsf{supp}}
\newcommand{\jlcs}{\mathcal{C\!S}}
\newcommand{\jlq}[1]{\ulcorner #1 \urcorner}
\newcommand{\jlimp}{\to}
\newcommand{\jlvd}{\vdash_{\!\mathcal{J}}}
\newcommand{\jlWhy}{\mathsf{Why}}
\newcommand{\jlPos}{\mathsf{PosBool}}

\newcommand{\jlprov}{\mathsf{prov}}
\newcommand{\jlfv}{\mathsf{fully\_verified}}
\newcommand{\jlsw}{\mathsf{survives\_without}}

\newcommand{\Sign}{\mathbf{Sign}}
\newcommand{\Sen}{\mathsf{Sen}}
\newcommand{\Mod}{\mathsf{Mod}}
\newcommand{\Th}{\mathbf{Th}}
\newcommand{\Thz}{\mathbf{Th}_0}
\newcommand{\instThl}{\mathbf{Th}_{\mathrm{list}}}
\newcommand{\SetC}{\mathbf{Set}}
\newcommand{\Cat}{\mathbf{Cat}}
\newcommand{\Ent}{\mathbf{Ent}}
\newcommand{\instPC}{\mathbf{PC}}
\newcommand{\piIns}{\pi\text{-}\mathbf{Ins}}
\newcommand{\Ins}{\mathbf{Ins}}
\newcommand{\instco}{\mathrm{co}}
\newcommand{\instP}{\mathcal{P}}
\newcommand{\instK}{\mathcal{K}}

\newcommand{\instJ}{\mathcal{J}}
\newcommand{\instJTh}{\mathcal{J}_{\mathrm{th}}}
\newcommand{\instI}{\mathcal{I}}
\newcommand{\instG}{\mathcal{G}}
\newcommand{\instF}{\mathcal{F}}
\newcommand{\instU}{\mathcal{U}}
\newcommand{\proofs}{\mathit{pf}}
\newcommand{\instPr}{\mathit{Pr}}

\newcommand{\instclo}[1]{#1^{\bullet}}
\newcommand{\insthat}[1]{\widehat{#1}}
\newcommand{\instax}[1]{\mathsf{ax}_{#1}}
\newcommand{\instimp}{\Rightarrow}

\newcommand{\Entity}{\mathsf{Entity}}

\newcommand{\RC}[1]{R_{#1}}
\newcommand{\rtype}[1]{\left[\,#1\,\right]}
\newcommand{\kgrcd}[1]{\left\{\,#1\,\right\}}
\newcommand{\ttrmeet}{\wedge}
\newcommand{\ttrjoin}{\vee}
\newcommand{\sing}[1]{\mathsf{Sing}_{#1}}
\newcommand{\sqsubclass}{\sqsubseteq_0}
\newcommand{\kganc}{\sqsubseteq^{*}}
\newcommand{\kgeqv}{\equiv_{\mathrm{cls}}}
\newcommand{\isa}{\mathsf{is\_a}}
\newcommand{\subclassof}{\mathsf{subclass\_of}}
\newcommand{\kglabels}{\mathsf{dom}}
\newcommand{\kgtel}{\mathsf{tel}}
\newcommand{\kgcanon}{\mathsf{canon}}
\newcommand{\kgReq}{\mathsf{Req}}
\newcommand{\kgReqz}{\mathsf{Req}_0}
\newcommand{\kgty}[1]{\tau(#1)}

\newcommand{\kgPhi}{\Phi_{\mathcal{G}}}

\begin{document}

\title{Typed Knowledge over Constructive Institutions: A Proof-Relevant Type Theory for Verified Science}

\author{Hans-Martin Will}
\affiliation{%
  \institution{The Eigenius Project}
  \country{United States}}
\email{hwill@acm.org}

\renewcommand{\shortauthors}{H.-M. Will}

\begin{abstract}
\noindent
Proof-assistant kernels with strict propositions---Lean~4, Agda's $\Prop$, and Coq/Rocq's $\mathsf{SProp}$---build proof irrelevance into conversion ($p \defeq q$ for $p, q : P$), equating all derivations of a theorem. While natural for pure mathematics, judgmental proof irrelevance breaks provenance in verified scientific stores, where predicates, refinements, and institutional maps must distinguish independent derivations for consilience and counterfactual retraction; with subsingleton elimination, it also breaks normalization and algorithmic transitivity.

We present the type-theoretic and institutional foundations of \emph{Eigenius}, a kernel for provenance-preserving scientific stores. First, we formulate a dependent kernel with cumulative universes ($\Prop = \Srt\,0$, $\Setu = \Srt\,1$, $\Typeu_k = \Srt\,(k+1)$ for $k \ge 1$), impredicative $\Prop$, inductive families with index-forced subsingleton elimination, and intensional identity types: conversion $\defeq$ is untyped orthogonal-CRS joinability, while proof irrelevance is a witnessed axiom ($\irrel : \Pi P{:}\Prop.\,\Pi p,q{:}P.\,\Id\,P\,p\,q$). Second, coercion-free subsumption unifies cumulativity, $\Pi$-covariant subtyping, and ontology inclusions; its bidirectional fragments ($\mathcal{K}_0, \mathcal{K}_1$) satisfy subject reduction, match declarative typing, and compile to heuristic-free checkers ($\mathcal{A}_0, \mathcal{A}_1$) sound and complete relative to strong normalization. Third, we link external verifiers via constructive institutions (\emph{Adopt}, \emph{Discharge}, \emph{Relativize}), internalize justification logic with Why-provenance $\supp(j,P)$, and type open-world knowledge graphs via TTR. Finally, we prove consistency in Grothendieck universes and mechanize the metatheory in Lean~4 with zero \code{sorry}.
\end{abstract}

\maketitle

\section{Introduction \& Epistemic Motivation}
\label{sec:intro}

Software agents and automated discovery pipelines increasingly author, execute, and synthesize the scientific record across computational biology, materials science, and clinical pharmacology. Yet the epistemic infrastructure of science remains split between two incompatible paradigms. On one side, semantic-web knowledge graphs (RDF, OWL, PROV-O)~\cite{provdm} and relational provenance systems~\cite{green-karvounarakis-tannen2007} track attribution and query lineage over open-world data, but lack a proof-checking kernel: an assertion that a conclusion follows from a dataset has the same syntactic standing as a hallucinated claim. On the other side, interactive theorem provers based on dependent type theory---such as Coq/Rocq~\cite{coq-refman}, Agda~\cite{agda}, and Lean~4~\cite{lean4}---provide small trusted kernels where claim admission is reduced to decidable type checking under the Brouwer--Heyting--Kolmogorov (BHK) propositions-as-types correspondence.

Bridging this divide requires turning a scientific repository into a typed knowledge store whose kernel checks deductive, statistical-transport, and schema invariants the way a compiler checks a program~\cite{will-brown-fuchs2026-theory,will-brown-fuchs2026}---not claiming that wet-lab assays or modeling choices are infallible, which instead enter as explicitly typed, graded ($\jdecl, \jobs$), and retractable premises. Yet building such a store directly on existing proof-assistant kernels exposes a fundamental tension between mathematical proof erasure and scientific provenance.

\subsection{Proof Assistants vs.\ Verified Knowledge Stores}
\label{sec:intro-pa-vs-ks}

In a mathematical proof assistant, the purpose of checking a proof term $p$ against a proposition $P : \Prop$ is solely to verify that $P$ is inhabited relative to a small, fixed list of foundational axioms (e.g., propositional, functional, and quotient extensionality plus classical choice in Lean~4). Once the kernel accepts $\vdash p : P$, the specific derivation $p$ carries no further epistemic load: any two proofs $p, q : P$ establish the exact same mathematical truth under the exact same axioms. To simplify programming with subset types $\{x : A \mid P\,x\}$ and proof-indexed data, Lean~4~\cite{carneiro2019,lean4}, Agda's $\Prop$, and Coq/Rocq's strict $\mathsf{SProp}$~\cite{gilbert2019} bake \emph{definitional proof irrelevance} directly into conversion (whereas Coq's default $\Prop$ keeps conversion untyped, embeds $\Prop \subty \Setu$ cumulatively, and postulates proof irrelevance only as an optional library axiom):
\[
  \inferrule*[right=Def-PI]
    {\ctx \vdash P : \Prop \\ \ctx \vdash p : P \\ \ctx \vdash q : P}
    {\ctx \vdash p \defeq q : P}
\]
Under \tttrule{Def-PI}, any two well-typed inhabitants of $P$ are definitionally equal during type checking, and extraction erases proofs entirely.

In a verified scientific knowledge store, by contrast, the leaves of an argument are an open-ended, evolving collection of empirical observations (assay readouts, sequencing matrices), methodological conventions (biomarker thresholds, domain bridges), and external computational verdicts (empirical-Bayes models, mixed-effects regressions, SMT certificates). In this setting, identifying proofs by conversion breaks three core operations:
\begin{enumerate}
\item \textbf{Multi-platform consilience.} A scientific claim $P : \Prop$ established independently by two orthogonal experimental modalities---such as a CRISPR--Cas9 knockout screen $p_1 : P$ and an RNA interference (shRNA) knockdown screen $p_2 : P$---has qualitatively stronger epistemic standing than a claim established by running a single screen twice. Under \tttrule{Def-PI}, $p_1 \defeq p_2$, so any predicate, record, or tuple $\tttpair{p_1}{p_2}$ recording the two derivations is definitionally equal to the redundant pair $\tttpair{p_1}{p_1}$.
\item \textbf{Counterfactual retraction and in-language audit.} Scientific premises are routinely revised or retracted when a cell line is cross-contaminated, an shRNA reagent exhibits seed-region off-target toxicity, or a statistical plan suffers from pseudoreplication. Although an external metaprogram can still scan raw declaration ASTs in the environment, \emph{inside the type theory} \tttrule{Def-PI} forces $F\,p_1 \defeq F\,p_2$ for every family $F$, inhabits $\Id\,P\,p_1\,p_2$ by $\refl\,p_2$, and equates downstream proofs through a retracted $p_1$ with those through a valid $p_2$.
\item \textbf{Normalization and transitivity of conversion.} Even inside the checker, \tttrule{Def-PI} makes conversion depend on typing (inferring types to test $\Prop$-inhabitation). Combining \tttrule{Def-PI} with impredicative $\Prop$ and large $\Id$-elimination breaks closed normalization~\cite{abel-coquand2020}, while large $\tttAcc$-elimination breaks strong normalization on open proofs~\cite{gilbert2019} and WHNF transitivity~\cite{carneiro2019}.
\end{enumerate}

At the same time, a scientific type theory \emph{cannot} abandon proof irrelevance altogether (e.g., replacing $\Prop$ by predicative $\Setu$, i.e., MLTT's $\mathcal{U}_0$): impredicative quantification over propositions is needed for domain closures and subtype invariants, and $\Setu$ computations must not branch on deductive proof syntax lest subset witnesses fail to interoperate. We therefore require a type theory where (i)~proof terms remain syntactically and definitionally distinct ($p_1 \not\defeq p_2$) so the store and institutional translations can inspect, transport, and audit them, while (ii)~all proofs of $P : \Prop$ are \emph{propositionally} equal via $\irrel\,P\,p_1\,p_2 : \Id\,P\,p_1\,p_2$, and data types cannot eliminate proof structure.

\subsection{Motivating Example: The WRN Synthetic Lethality Claim}
\label{sec:intro-wrn}

To make this tension concrete, consider our end-to-end formalization~\cite{will-kasarskis2026-wrn} (machine-checking all 55 statistical and domain verdicts across 10 layers) of the landmark oncology study by \citet{chan2019}, which identified the RecQ-family DNA helicase \emph{WRN} as a synthetic lethal dependency in microsatellite-unstable ($\mathrm{MSI}$) cancers. Let $\Entity : \Setu$ be the type of repository resources, with declared ontology classes $\mathsf{Gene}, \mathsf{Cohort} \subty \Entity$, constants $\mathsf{WRN} : \mathsf{Gene}$ and $\mathsf{MSI}, \mathsf{MSS} : \mathsf{Cohort}$, and target discovery proposition
\[
  \Phi_{\mathrm{WRN}} \;:=\; \mathsf{SelectivelyEssential}(\mathsf{WRN}, \mathsf{MSI}) \;:\; \Prop.
\]
In the study, $\Phi_{\mathrm{WRN}}$ is established by two independent genome-wide functional genomics screens (Figure~\ref{fig:wrn-motivating}):
\begin{itemize}
\item \textbf{CRISPR--Cas9 knockout derivation ($p_{\mathrm{CRISPR}}$).} Over the Project Achilles dataset $s_{\mathrm{Ach}}$ (CERES dependency scores across 17,634 genes in $37$ $\mathrm{MSI}$ vs.\ $91$ $\mathrm{MSS}$ cell lines), an empirical-Bayes moderated-$t$ pipeline (\code{dd\_achilles\_limma}) verifies that WRN ranks \#1 genome-wide with false-discovery rate $Q = 4.81 \times 10^{-24} < 10^{-20}$. Via a statistical institution comorphism (\S\ref{sec:inst-env}), this emits a checked witness $w_{\mathrm{Ach}} : \mathsf{ModeratedT}(s_{\mathrm{Ach}}, \mathsf{WRN}) < -10$, which composes with the declared domain bridge $b_{\mathrm{CRISPR}} : \mathsf{ModeratedT}(s_{\mathrm{Ach}}, \mathsf{WRN}) < -10 \to \Phi_{\mathrm{WRN}}$ (\emph{Adopt}, tagged $\jdecl$ in \S\ref{sec:inst-jl}) to form $p_{\mathrm{CRISPR}} := b_{\mathrm{CRISPR}}\,w_{\mathrm{Ach}} : \Phi_{\mathrm{WRN}}$.
\item \textbf{RNAi shRNA knockdown derivation ($p_{\mathrm{RNAi}}$).} Over the independent Project DRIVE dataset $s_{\mathrm{DRV}}$ (DEMETER2 shRNA scores across $32$ $\mathrm{MSI}$ vs.\ $413$ $\mathrm{MSS}$ lines), the corresponding analysis (\code{dd\_drive\_limma}) yields WRN at rank \#1 with $Q = 1.46 \times 10^{-45}$, producing a witness $w_{\mathrm{DRV}} : \mathsf{ModeratedT}(s_{\mathrm{DRV}}, \mathsf{WRN}) < -10$ and, via declared bridge $b_{\mathrm{RNAi}}$, a second kernel proof term $p_{\mathrm{RNAi}} := b_{\mathrm{RNAi}}\,w_{\mathrm{DRV}} : \Phi_{\mathrm{WRN}}$.
\end{itemize}

\begin{figure}[t]
\centering
\resizebox{\columnwidth}{!}{%
\begin{tikzpicture}[
  font=\scriptsize,
  >={Latex[length=3.8pt]},
  box/.style={draw=black!65, rounded corners=2pt, inner sep=3.5pt, align=center},
  obsbox/.style={box, fill=blue!8},
  decbox/.style={box, fill=purple!8},
  prfbox/.style={box, fill=green!10},
  goalbox/.style={box, fill=green!18, line width=0.7pt}
]
\node[obsbox] (ach) at (0, 1.35) {Achilles CRISPR screen\\$w_{\mathrm{Ach}} : \mathsf{ModT}(s_{\mathrm{Ach}}) < -10$};
\node[decbox] (b1) at (3.1, 1.35) {Bridge $b_{\mathrm{CRISPR}}$\\$\mathsf{ModT}(s_{\mathrm{Ach}}) \to \Phi_{\mathrm{WRN}}$};
\node[prfbox] (p1) at (5.9, 1.35) {$p_{\mathrm{CRISPR}} := b_{\mathrm{CRISPR}}\,w_{\mathrm{Ach}}$\\$: \Phi_{\mathrm{WRN}}$};

\node[obsbox] (drv) at (0, 0) {DRIVE shRNA screen\\$w_{\mathrm{DRV}} : \mathsf{ModT}(s_{\mathrm{DRV}}) < -10$};
\node[decbox] (b2) at (3.1, 0) {Bridge $b_{\mathrm{RNAi}}$\\$\mathsf{ModT}(s_{\mathrm{DRV}}) \to \Phi_{\mathrm{WRN}}$};
\node[prfbox] (p2) at (5.9, 0) {$p_{\mathrm{RNAi}} := b_{\mathrm{RNAi}}\,w_{\mathrm{DRV}}$\\$: \Phi_{\mathrm{WRN}}$};

\draw[->] (ach) -- (b1);
\draw[->] (b1) -- (p1);
\draw[->] (drv) -- (b2);
\draw[->] (b2) -- (p2);

\draw[<->, dashed, thick, red!75!black] (p1.south) -- (p2.north)
  node[midway, left=3pt, align=right, font=\tiny] {$p_{\mathrm{CRISPR}} \not\defeq p_{\mathrm{RNAi}}$\\$\irrel\,\Phi_{\mathrm{WRN}}\,p_{\mathrm{CRISPR}}\,p_{\mathrm{RNAi}}$};
\end{tikzpicture}%
}
\vspace{-4pt}
\caption{Two independent experimental derivations $p_{\mathrm{CRISPR}}, p_{\mathrm{RNAi}} : \Phi_{\mathrm{WRN}}$ in the WRN synthetic lethality study~\cite{chan2019,will-kasarskis2026-wrn}. Eigenius keeps $p_{\mathrm{CRISPR}} \not\defeq p_{\mathrm{RNAi}}$ distinct under conversion while witnessing their propositional equality via $\irrel$.}
\Description{Two parallel derivation paths from Achilles CRISPR screen and DRIVE shRNA screen to proofs of WRN selective essentiality, remaining distinct under conversion while connected by propositional proof irrelevance.}
\label{fig:wrn-motivating}
\vspace{-6pt}
\end{figure}

A second, equally revealing instance occurs in the wet-lab validation arm of \citet{chan2019}: in the KM12 viability competition assay ($\Psi_{\mathrm{KM12}} : \Prop$), our replication~\cite{will-kasarskis2026-wrn} uncovered that the published two-way ANOVA ($p_{\mathrm{tech}} : \Psi_{\mathrm{KM12}}$, reporting $P = 2.74 \times 10^{-19}$ on $25$ residual degrees of freedom) treats technical replicates within the same sgRNA guide as independent biological units---a classic pseudoreplication error. Re-analyzing the pinned data with a linear mixed-effects model that treats guide as a random effect (\code{lmer(value \textasciitilde\ is\_WRN + (1|guide))}) yields a second, methodologically sound derivation $p_{\mathrm{bio}} : \Psi_{\mathrm{KM12}}$ with $P = 2.15 \times 10^{-6}$---thirteen orders of magnitude weaker, though still significant.

\paragraph{Why judgmental proof irrelevance fails.}
If the kernel enforced \tttrule{Def-PI}, then $p_{\mathrm{CRISPR}} \defeq p_{\mathrm{RNAi}}$ and $p_{\mathrm{tech}} \defeq p_{\mathrm{bio}}$. While external tooling can still inspect raw stored declarations, \emph{in-language} typing collapses them: for $f : \Phi_{\mathrm{WRN}} \to \Psi_{\mathrm{KM12}} \to \chi_{\mathrm{SL}}$,
\[
  f\;p_{\mathrm{CRISPR}}\;p_{\mathrm{tech}} \;\defeq\; f\;p_{\mathrm{RNAi}}\;p_{\mathrm{bio}},
\]
$\refl\,p_{\mathrm{bio}}$ inhabits $\Id\,\Psi_{\mathrm{KM12}}\,p_{\mathrm{tech}}\,p_{\mathrm{bio}}$ (so $\{q : \Psi_{\mathrm{KM12}} \mid \Id\,\Psi_{\mathrm{KM12}}\,q\,p_{\mathrm{bio}}\}$ accepts the retracted $p_{\mathrm{tech}}$), and any conversion-quotiented institutional translation collapses $\proofs_{\instK}(T)$ to $\mathit{ent}(\instK)^\natural$ (Theorem~\ref{thm:kernel-pc}(2)).

\paragraph{How Eigenius resolves it.}
Eigenius separates three levels that existing systems conflate:
\begin{enumerate}
\item \emph{Intensional conversion on $\Prop$.} Definitional equality $\defeq$ in the Eigenius kernel is untyped joinability under an orthogonal rewriting system (\S\ref{sec:kernel-rewriting}). Because no rewrite rule equates arbitrary inhabitants of $\Prop$, $p_{\mathrm{CRISPR}} \not\defeq p_{\mathrm{RNAi}}$ and $p_{\mathrm{tech}} \not\defeq p_{\mathrm{bio}}$ remain distinct normal forms in the witness set $\proofs_{\instK}(T)$ of the kernel's proof calculus (\S\ref{sec:inst-env}).
\item \emph{Witnessed propositional proof irrelevance.} The global environment declares the axiom $\irrel : \Pi P{:}\Prop.\,\Pi p{:}P.\,\Pi q{:}P.\,\Id\,P\,p\,q$ (\S\ref{sec:kernel-pi}). Whenever a proof-irrelevant data structure needs to transport along two proofs of $\Phi_{\mathrm{WRN}}$, the term $\irrel\,\Phi_{\mathrm{WRN}}\,p_{\mathrm{CRISPR}}\,p_{\mathrm{RNAi}}$ witnesses their propositional equality, while the resulting transport term $\Jelim\,p_{\mathrm{CRISPR}}\,C\,d\,p_{\mathrm{RNAi}}\,(\irrel\,\Phi_{\mathrm{WRN}}\,p_{\mathrm{CRISPR}}\,p_{\mathrm{RNAi}})$ remains a neutral term recording the exact bridge.
\item \emph{Internalized justification terms and provenance support.} Alongside deductive proofs in $\Prop$, the empirical and methodological grounds of $\Phi_{\mathrm{WRN}}$ are represented as inductive data terms in $\JTerm : \Setu$ (\S\ref{sec:inst-jl}), namely $j_{\mathrm{WRN}} := \jsum(\japp(\jdecl(b_{\mathrm{CRISPR}}),\allowbreak \jobs(w_{\mathrm{Ach}})),\allowbreak \japp(\jdecl(b_{\mathrm{RNAi}}),\allowbreak \jobs(w_{\mathrm{DRV}})))$. The kernel's computable support function evaluates its proposition-relative support in the Why-provenance semiring $\jlWhy(\jlLeaf)$:
\[
  \supp(j_{\mathrm{WRN}}, \Phi_{\mathrm{WRN}}) = \bigl\{\{b_{\mathrm{CRISPR}}, w_{\mathrm{Ach}}\},\; \{b_{\mathrm{RNAi}}, w_{\mathrm{DRV}}\}\bigr\}.
\]
Retracting $b_{\mathrm{RNAi}}$ leaves the first alternative intact, so the query $\jlsw(j_{\mathrm{WRN}},\allowbreak \Phi_{\mathrm{WRN}},\allowbreak b_{\mathrm{RNAi}})$ evaluates to $\mathsf{true}$ in $O(|j|^3)$ semiring operations over the stored justification term (Theorem~\ref{thm:jl-semiring}(2)) without re-running external containers.
\end{enumerate}

\subsection{Contributions \& Road Map}
\label{sec:intro-contributions}

Our \emph{primary} contribution (\S\S\ref{sec:kernel}, \ref{sec:subsumption}, \ref{sec:semantics}) is a proof-relevant dependent kernel with coercion-free cumulative/ontology subsumption and a normalization-free Grothendieck consistency proof; \S\ref{sec:institutions} supplies the minimal institutional, justification, and TTR layers required to deploy it over multi-verifier stores. Full proofs appear in our 249-page monograph~\cite{will-brown-fuchs2026-theory} (\url{https://doi.org/10.5281/zenodo.23964990}), alongside a tutorial primer and reference checker~\cite{will_2026_24211920} (\url{https://doi.org/10.5281/zenodo.24211920}) and our Lean~4 mechanization (\code{EigeniusTheory}, ${\sim}58{,}800$ lines, 119 files, zero \code{sorry} at \url{https://github.com/eigenius/theory}):
\begin{itemize}
\item \textbf{Eigenius Kernel \& Orthogonal Conversion (\S\ref{sec:kernel}).} We formulate a dependent calculus with cumulative universes, impredicative $\Prop$, strictly positive inductive families with index-forced subsingleton elimination, and intensional identity types. Defining conversion $\defeq$ via an untyped regular CRS with independent linear pattern variables yields Klop confluence, level-equivalence stability, type-former injectivity, and avoidance of the Abel--Coquand normalization failure, while internalizing proof irrelevance via $\irrel$.
\item \textbf{Unified Subsumption, Bidirectional Typing \& \emph{Unit}-Style Checking (\S\ref{sec:subsumption}).} We combine universe cumulativity ($\Prop \not\subty \Setu \subty \Typeu_1 \subty \cdots$), codomain-covariant $\Pi$-subtyping with convertible domains, and ontology class inclusions ($\kganc$) in a single coercion-free judgment $\ctx \vdash A \subty B$, prove transitivity admissibility, subject reduction under annotated reduction $\tttstep_a$, and declarative soundness and completeness for $\mathcal{K}_0$ and $\mathcal{K}_1$, and compile them to heuristic-free WHNF and untyped-NbE checkers (sound and complete relative to strong normalization) in the minimalist spirit of Coquand's \emph{Mini-TT} and \emph{Unit}~\cite{coquand2026unit,minitt}.
\item \textbf{Constructive Institutions, Justification \& Knowledge Graphs (\S\ref{sec:institutions}).} We organize kernel theories and external verifiers into a three-level institutional hierarchy (Meseguer proof calculi $\instP$, $\pi$-institutions, and Mariano--Pinto institutions $\instG(\instJ)$), prove that $\irrel$-quotienting collapses $\instK$ to $\mathit{ent}(\instK)^\natural$, classify comorphisms under \emph{Adopt}, \emph{Discharge}, and \emph{Relativize}, internalize a Curry-style justification calculus $\JTerm$ with $O(n^2)$ evidence checking and $O(n^3)$ semiring provenance queries, and embed Cooper's TTR for open-world coinductive knowledge graphs.
\item \textbf{Stratified Set-Theoretic Semantics \& Mechanization (\S\ref{sec:semantics}).} We construct a direct stratified model in Grothendieck universes $V_0 \in V_1 \in \cdots$ where sorts are read off bidirectional derivations, proofs denote $\emptyset$, data functions denote domain-recording graphs, and subsumption denotes set inclusion $\llbracket A \rrbracket \subseteq \llbracket B \rrbracket$, establishing consistency for $\mathcal{K}_0$, $\mathcal{K}_1$, the declarative kernel, and the knowledge-graph calculus (discharged unconditionally in Lean~4 for finite universe bounds). Section~\ref{sec:related} contrasts Eigenius with $\mathsf{SProp}$, OTT, minimalist checkers (\emph{Mini-TT}, \emph{Unit}), coercive subtyping, Justification Logic, and MMT/Hets.
\end{itemize}

\section{The Eigenius Kernel \& Orthogonal Conversion}
\label{sec:kernel}

The Eigenius kernel combines the dependent function, pair, and intensional identity types of Martin-L\"of's intuitionistic type theory~\cite{martinlof1984} with the cumulative universe hierarchy of Luo's Extended Calculus of Constructions (ECC)~\cite{luo1994} and the impredicative $\Prop$ and inductive families of the Calculus of Inductive Constructions (CIC)~\cite{coquand-paulin1990,werner1997}, while keeping definitional equality entirely type-free.

\subsection{Syntax \& Universe Hierarchy}
\label{sec:kernel-syntax}

\paragraph{Universe levels and cumulativity.}
Let $u, v$ range over universe-level parameters. The grammar of \emph{universe level expressions} is:
\[
  \ell \;::=\; \lzero \;\mid\; \lsucc\,\ell \;\mid\; \lmax\,\ell_1\,\ell_2 \;\mid\; \limax\,\ell_1\,\ell_2 \;\mid\; u.
\]
A valuation $v : \mathrm{Param} \to \mathbb{N}$ interprets $\lzero$ as $0$, $\lsucc\,\ell$ as $\llbracket\ell\rrbracket_v + 1$, $\lmax\,\ell_1\,\ell_2$ as $\max(\llbracket\ell_1\rrbracket_v, \llbracket\ell_2\rrbracket_v)$, and the \emph{impredicative maximum} $\limax\,\ell_1\,\ell_2$ as $\llbracket\limax\,\ell_1\,\ell_2\rrbracket_v = 0$ if $\llbracket\ell_2\rrbracket_v = 0$ and $\max(\llbracket\ell_1\rrbracket_v, \llbracket\ell_2\rrbracket_v)$ otherwise.
Two level expressions are \emph{equivalent}, written $\ell_1 \tttsim \ell_2$, if $\llbracket\ell_1\rrbracket_v = \llbracket\ell_2\rrbracket_v$ for all valuations $v$; $\ell_1 \le \ell_2$ holds if $\llbracket\ell_1\rrbracket_v \le \llbracket\ell_2\rrbracket_v$ for all $v$. The \emph{cumulative level order} $m \cumle n$ holds if $m \le n$ and $(\llbracket m\rrbracket_v = 0 \iff \llbracket n\rrbracket_v = 0)$ for all $v$. Thus $\lzero \cumle n$ iff $n \tttsim \lzero$: the universe of propositions $\Prop := \Srt\,\lzero$ is isolated from the predicative data universes $\Setu := \Srt\,1$ and $\Typeu_k := \Srt\,(k+1)$ ($k \ge 1$), which form a cumulative chain $\Setu \subty \Typeu_1 \subty \Typeu_2 \subty \cdots$ with $\Prop : \Setu$ and $\Srt\,\ell : \Srt\,(\lsucc\,\ell)$, but $\Prop \not\subty \Setu$.

\begin{proposition}[Decidability of level relations]\label{prop:lvl-decidable}
For level expressions $\ell_1, \ell_2$ with $k$ free parameters and maximum $\lsucc$-nesting depth $d$, $\ell_1 \le \ell_2$ (and hence $\ell_1 \tttsim \ell_2$ and $\ell_1 \cumle \ell_2$) is decidable by case-splitting on the $2^k$ zero/positive sign substitutions or by testing all $(d+3)^k$ valuations in $\{0, 1, \ldots, d+2\}^k$.
\end{proposition}
\begin{proof}
Substituting $u_r \mapsto \lzero$ or $u_r \mapsto \lsucc\,u'_r$ eliminates every $\limax$, normalizing each side to $\max(\{c\} \cup \{u'_i + 1 + a_i\})$ with $c, a_i \le d$; any failure of $\ell_1 \le \ell_2$ on a branch $u'_i + 1 + a_i$ is witnessed at $u'_r = 0$ ($r \neq i$) and $u'_i = d + 1$, i.e., $v \in \{0, 1, \ldots, d+2\}^k$.
\end{proof}

\paragraph{Terms, contexts, and declarations.}
The syntax of terms is:
\begin{align*}
  a, b, A, B \;::=\;& x \;\mid\; \Srt\,\ell \;\mid\; \tttinst{c}{\bar\ell} \;\mid\; \Pi x{:}A.\,B \;\mid\; \tttlam{x}{A} b \;\mid\; a\,b \\
  &\mid\; \Sigma x{:}A.\,B \;\mid\; \tttpair{a}{b} \;\mid\; \fst\,a \;\mid\; \snd\,a \\
  &\mid\; \Id\,A\,a\,b \;\mid\; \refl\,a \;\mid\; \Jelim\,a\,C\,d\,b\,p \;\mid\; \tttann{a}{A}
\end{align*}
Here $\tttinst{c}{\bar\ell}$ instantiates a global constant $c$ at universe levels $\bar\ell$, every $\lambda$-abstraction $\tttlam{x}{A} b$ carries its domain type $A$, and $\tttann{a}{A}$ is an explicit type annotation used by the bidirectional kernel (\S\ref{sec:subsumption-bidir}). The erasure $|t|$ strips all $\tttann{\cdot}{\cdot}$ nodes recursively.

A \emph{global environment} $\tttEnv$ is a well-formed sequence of declarations ($\tttEnv \tttwf$), each checked in the prefix preceding it and parameterized by level variables $\bar u$:
(1)~an \emph{axiom} $\mathsf{axiom}\ \tttinst{c}{\bar u} : A$;
(2)~a \emph{definition} $\mathsf{def}\ \tttinst{c}{\bar u} : A := t$ (marked either \emph{transparent} or \emph{opaque}); or
(3)~a single \emph{inductive family} $\mathsf{inductive}\ \tttinst{D}{\bar u}\ (\Delta_P) : \Pi\Delta_I.\,\Srt\,\ell$ with uniform parameters $\Delta_P$, indices $\Delta_I$, and constructors $c_k : \Pi\Delta_P.\,\Pi\Delta_k.\,D\,\bar p\,\bar t_k$ (mutual/nested families and primitive quotients are excluded from the kernel).

\paragraph{Formation rules and subsingleton elimination.}
Figure~\ref{fig:kernel-typing} gives the core declarative rules for universes, $\Pi$-types, $\Sigma$-types, and identity types. By Rule~\tttrule{$\Pi$-F}, $\Pi x{:}A.\,B$ has sort $\Srt\,(\limax\,\ell_1\,\ell_2)$, so $\Pi x{:}A.\,P : \Prop$ whenever $P : \Prop$ regardless of $\ell_1$ (impredicativity of $\Prop$). By contrast, Rule~\tttrule{$\Sigma$-F} uses $\lmax\,\ell_1\,\ell_2$: a pair over $A : \Setu$ lives in $\Setu$ (strong $\Sigma$ with $\fst, \snd$), whereas $\exists x{:}A.\,P$ is an inductive family $\tttExists\,A\,(\tttlam{x}{A} P) : \Prop$ (allowing $\fst$ out of $(\exists b{:}\tttBool.\,\top) : \Prop$ under $\irrel$ would equate $\mathsf{true} = \mathsf{false}$). Identity types $\Id\,A\,a\,b$ inhabit $\Prop$ (\tttrule{Id-F}), while $\Jelim\,a\,C\,d\,b\,p$ eliminates into any $\Srt\,\ell'$ (\tttrule{Id-E}).

\begin{figure}[t]
\footnotesize
\begin{mathpar}
\inferrule*[right=\tttrule{Sort}]
  {\ctx \tttctx}
  {\ctx \vdash \Srt\,\ell : \Srt\,(\lsucc\,\ell)}
\and
\inferrule*[right=\tttrule{$\Pi$-F}]
  {\ctx \vdash A : \Srt\,\ell_1 \\ \ctx, x{:}A \vdash B : \Srt\,\ell_2}
  {\ctx \vdash \Pi x{:}A.\,B : \Srt\,(\limax\,\ell_1\,\ell_2)}
\\
\inferrule*[right=\tttrule{$\Pi$-I}]
  {\ctx, x{:}A \vdash b : B}
  {\ctx \vdash \tttlam{x}{A} b : \Pi x{:}A.\,B}
\and
\inferrule*[right=\tttrule{$\Pi$-E}]
  {\ctx \vdash f : \Pi x{:}A.\,B \\ \ctx \vdash a : A}
  {\ctx \vdash f\,a : B[a/x]}
\\
\inferrule*[right=\tttrule{$\Sigma$-F}]
  {\ctx \vdash A : \Srt\,\ell_1 \\ \ctx, x{:}A \vdash B : \Srt\,\ell_2}
  {\ctx \vdash \Sigma x{:}A.\,B : \Srt\,(\lmax\,\ell_1\,\ell_2)}
\and
\inferrule*[right=\tttrule{$\Sigma$-I}]
  {\ctx \vdash a : A \\ \ctx \vdash b : B[a/x] \\ \ctx, x{:}A \vdash B : \Srt\,\ell}
  {\ctx \vdash \tttpair{a}{b} : \Sigma x{:}A.\,B}
\\
\inferrule*[right=\tttrule{Id-F}]
  {\ctx \vdash A : \Srt\,\ell \\ \ctx \vdash a : A \\ \ctx \vdash b : A}
  {\ctx \vdash \Id\,A\,a\,b : \Prop}
\and
\inferrule*[right=\tttrule{Id-I}]
  {\ctx \vdash a : A}
  {\ctx \vdash \refl\,a : \Id\,A\,a\,a}
\\
\inferrule*[right=\tttrule{Id-E}]
  {\ctx \vdash a : A \\ \ctx \vdash b : A \\ \ctx \vdash p : \Id\,A\,a\,b \\
   \ctx \vdash C : \Pi y{:}A.\,\Id\,A\,a\,y \to \Srt\,\ell' \\ \ctx \vdash d : C\,a\,(\refl\,a)}
  {\ctx \vdash \Jelim\,a\,C\,d\,b\,p : C\,b\,p}
\\
\inferrule*[right=\tttrule{Conv}]
  {\ctx \vdash a : A \\ \ctx \vdash B : \Srt\,\ell \\ A \defeq B}
  {\ctx \vdash a : B}
\and
\inferrule*[right=\tttrule{Sub}]
  {\ctx \vdash a : A \\ \ctx \vdash B : \Srt\,\ell \\ \ctx \vdash A \subty B}
  {\ctx \vdash a : B}
\end{mathpar}
\vspace{-8pt}
\caption{Core declarative typing rules of the Eigenius kernel (projections $\fst, \snd$, constants, and annotations omitted).}
\Description{Core declarative typing rules of the Eigenius kernel for sorts, dependent products, dependent sums, intensional identity types, conversion, and subsumption.}
\label{fig:kernel-typing}
\vspace{-6pt}
\end{figure}

For an inductive declaration $\mathsf{inductive}\ \tttinst{D}{\bar u}\ (\Delta_P) : \Pi\Delta_I.\,\Srt\,\ell$ to be \emph{admissible} in $\tttEnv$, each constructor argument $y_j{:}B_j$ must be either non-recursive ($D \notin \mathsf{FV}(B_j)$) or \emph{strictly positive} of the form $B_j = \Pi\Theta_j.\,D\,\bar p\,\bar s_j$, and when $\ell \not\tttsim \lzero$ (a data family), the sort $\ell_{k,j}$ of each $B_j$ must satisfy $\ell_{k,j} \le \ell$ (preventing Girard-style paradoxes). To preserve consistency with proof irrelevance in $\Prop$, the motive $C : \Pi\Delta_I.\,D\,\bar p\,\bar\imath \to \Srt\,v$ of the derived recursor $\rec_D$ may target an arbitrary universe $v$ (\emph{large elimination}) if either $\ell \not\tttsim \lzero$ or $D$ is a \emph{syntactic subsingleton}:

\begin{definition}[Subsingleton criterion and index forcing]\label{def:subsingleton}
An admissible inductive family $D$ with target sort $\ell \tttsim \lzero$ is a \emph{subsingleton} if it has at most one constructor $c_1 : \Pi\Delta_P.\,\Pi(y_1{:}B_1)\cdots(y_m{:}B_m).\,D\,\bar p\,\bar t_1$, and every constructor argument $y_j{:}B_j$ either:
\begin{enumerate}
\item is a \emph{proof} ($\Delta_P, y_1{:}B_1, \ldots, y_{j-1}{:}B_{j-1} \vdash B_j : \Prop$), or
\item is \emph{index-forced}: $y_j$ occurs as one of the result indices ($t_{1,i} = y_j$) and $B_j$ equals the $i$-th index type in $\Delta_I$ instantiated at $t_{1,1}, \ldots, t_{1,i-1}$ (up to erasure).
\end{enumerate}
Whenever~(2) holds for $y_j$, the $\iota$-rule for $\rec_D$ reads $y_j$ off the recursor's $i$-th index $\imath_i$ rather than field $a_j$.
\end{definition}

Adapting Lean's syntactic subsingleton criterion~\cite{carneiro2019,felicissimo2026acc}, Definition~\ref{def:subsingleton} admits $\tttFalse$, $\top$, $\mathsf{And}\,P\,Q$, $\mathsf{Eq}$, and $\tttAcc\,(A : \Srt\,\ell)\,(R : A \to A \to \Prop) : A \to \Prop$ (whose constructor $\mathsf{intro} : \Pi x{:}A.\,(\Pi y{:}A.\,R\,y\,x \to \tttAcc\,A\,R\,y) \to \tttAcc\,A\,R\,x$ forces $x : A$ at the result index), while rejecting $\tttOr$ and $\tttExists$ ($v := \lzero$). Two refinements distinguish Definition~\ref{def:subsingleton}: (i)~under universe cumulativity, $B_j$ must match the exact index type in $\Delta_I$ (if $D : \Typeu_1 \to \Prop$ had $\mathsf{mk} : \Pi X{:}\Setu.\,D\,X$, an index checked at $\Typeu_1$ could not instantiate $X : \Setu$), and (ii)~reading forced fields off $\bar\imath$ rather than $a_j$ keeps $\iota$ sound when proofs denote $\emptyset$ (\S\ref{sec:semantics}).

\subsection{Definitional Equality as Orthogonal Rewriting}
\label{sec:kernel-rewriting}

Unlike Coq's $\mathsf{SProp}$, Agda's $\mathsf{Prop}$, or Lean~4's kernel, conversion in Eigenius---as in pure Martin-L\"of checkers such as \emph{Mini-TT}~\cite{minitt} and \emph{Unit}~\cite{coquand2026unit}---consults no typing derivations: one-step reduction $\tttstep$ on raw terms is the compatible closure of the seven rules in Figure~\ref{fig:kernel-reduction}.

\begin{figure}[t]
\footnotesize
\begin{mathpar}
\inferrule*[right=\tttrule{$\Pi$-$\beta$}]
  { }
  {(\tttlam{x}{A} b)\,a \tttstep b[a/x]}
\and
\inferrule*[right=\tttrule{$\Sigma$-$\beta_1$}]
  { }
  {\fst\,\tttpair{a}{b} \tttstep a}
\and
\inferrule*[right=\tttrule{$\Sigma$-$\beta_2$}]
  { }
  {\snd\,\tttpair{a}{b} \tttstep b}
\\
\inferrule*[right=\tttrule{Ann-$\tttstep$}]
  { }
  {\tttann{a}{A} \tttstep a}
\and
\inferrule*[right=\tttrule{$\delta$}]
  {(\mathsf{def}\ \tttinst{c}{\bar u} : A := t) \in \tttEnv \text{ transparent}}
  {\tttinst{c}{\bar\ell} \tttstep t[\bar\ell/\bar u]}
\\
\inferrule*[right=\tttrule{Id-$\iota$}]
  { }
  {\Jelim\,a\,C\,d\,b\,(\refl\,a') \tttstep d}
\\
\inferrule*[right=\tttrule{$\iota$}]
  {c_k \text{ is the } k\text{-th constructor of } D \text{ in } \tttEnv}
  {\tttinst{\rec_D}{\bar\ell, \ell'}\,\bar p\,C\,\bar f\,\bar\imath\,(\tttinst{c_k}{\bar\ell'}\,\bar p'\,\bar a) \\
   \tttstep\; f_k\,\bar a^*\,\bigl(\lambda\Theta_j.\,\tttinst{\rec_D}{\bar\ell, \ell'}\,\bar p\,C\,\bar f\,(\bar s_j[\bar a^*/\bar y])\,(a_j\,\overline{\Theta_j})\bigr)_{j \in R_k}}
\end{mathpar}
\vspace{-8pt}
\caption{Untyped orthogonal contraction rules of the Eigenius kernel. In \tttrule{$\iota$}, $a_j^* := \imath_i$ if $y_j$ is forced by index $i$ (Definition~\ref{def:subsingleton}) and $a_j^* := a_j$ otherwise.}
\Description{Untyped orthogonal contraction rules of the Eigenius kernel.}
\label{fig:kernel-reduction}
\vspace{-6pt}
\end{figure}

Two design choices in Figure~\ref{fig:kernel-reduction} warrant emphasis:
\begin{enumerate}
\item In \tttrule{Id-$\iota$}, base point $a$, endpoint $b$, and argument $a'$ of $\refl\,a'$ are \emph{independent pattern variables}: the rule never tests $a = b = a'$. Requiring syntactic identity would break left-linearity, and $a, a'$ may step independently under $\tttstep$ before \tttrule{Id-$\iota$} fires; on well-typed redexes, typing guarantees $a \defeq a' \defeq b$.
\item Likewise, in \tttrule{$\iota$}, the recursor's parameters $\bar p$, levels $\bar\ell$, and indices $\bar\imath$ are pattern variables independent of the constructor's parameters $\bar p'$, levels $\bar\ell'$, and field instances $\bar t_k[\bar a/\bar y]$. Because \tttrule{Id-$\iota$} and \tttrule{$\iota$} match these positions without testing equality, subject reduction is proved for bidirectional \emph{annotated reduction} $\tttstep_a$ (Theorem~\ref{thm:bidir-metatheory}(1)) rather than unrestricted untyped $\tttstep$.
\end{enumerate}

Let $\sim_{\mathrm{lvl}}$ be the congruence on raw terms equating $\Srt\,\ell \sim_{\mathrm{lvl}} \Srt\,\ell'$ for $\ell \tttsim \ell'$ and $\tttinst{c}{\bar\ell} \sim_{\mathrm{lvl}} \tttinst{c}{\bar\ell'}$ for componentwise $\ell_i \tttsim \ell'_i$.

\begin{definition}[Definitional equality]\label{def:defeq}
Two terms $a$ and $b$ are \emph{definitionally equal} over $\tttEnv$, written $a \defeq b$ (or $a \defeq_K b$), if there exist reducts $a', b'$ such that $a \tttstep^* a'$, $b \tttstep^* b'$, and $a' \sim_{\mathrm{lvl}} b'$.
\end{definition}

\begin{theorem}[Confluence, transitivity, and type-former injectivity]\label{thm:confluence}
Over any well-formed environment $\tttEnv$:
\begin{enumerate}
\item One-step reduction $\tttstep$ is confluent, and $\tttstep^*$ commutes with level equivalence $\sim_{\mathrm{lvl}}$.
\item Definitional equality $\defeq$ is an equivalence relation (in particular, transitive) and a congruence with respect to every term former and substitution.
\item Every type former and abstraction is injective under $\defeq$: $\Pi x{:}A.\,B \defeq \Pi x{:}A'.\,B'$, $\Sigma x{:}A.\,B \defeq \Sigma x{:}A'.\,B'$, and $\tttlam{x}{A} B \defeq \tttlam{x}{A'} B'$ each imply $A \defeq A'$ and $B \defeq B'$; $\Id\,A\,a\,b \defeq \Id\,A'\,a'\,b'$ implies $A \defeq A', a \defeq a', b \defeq b'$; and $D\,\bar p\,\bar\imath \defeq D\,\bar p'\,\bar\imath'$ implies $\bar p \defeq \bar p', \bar\imath \defeq \bar\imath'$.
\item Types with distinct head formers ($\Srt, \Pi, \Sigma, \Id, D$), or universes $\Srt\,\ell, \Srt\,\ell'$ with $\ell \not\tttsim \ell'$, are never definitionally equal.
\end{enumerate}
\end{theorem}

\begin{proof}[Proof sketch]
Every contraction rule in Figure~\ref{fig:kernel-reduction} is left-linear (no pattern variable occurs twice on a left-hand side), and because each constant $c$ in $\tttEnv$ is either a transparent definition, a constructor, or a recursor (never two at once), no two left-hand sides overlap at the root or at any non-variable subterm position. Thus $\tttstep$ is an orthogonal (regular) combinatory reduction system (CRS), which is confluent by Klop's theorem~\cite[Theorem~II.3.11]{klop1980}. Because $\delta$ and $\iota$ substitute level expressions homomorphically and level equivalence $\tttsim$ is a congruence under level substitution (\code{RuleSet.LevelCompat} in our Lean~4 formalization), $\sim_{\mathrm{lvl}}$ commutes with multi-step reduction (\code{RedStar.lequiv}), yielding~(1) and~(2). Finally, no contraction rule has $\Srt, \Pi, \Sigma, \Id$, an applied inductive type former $D$, or $\lambda$ at the head of its left-hand side, so reduction preserves head formers and reduces subterms independently, giving~(3) and~(4) (\code{DefEq.pi\_inj}, \code{sigma\_inj}, \code{ident\_inj}, \code{ne\_of\_head\_ne} in \code{TypeTheory/Confluence.lean}, \code{DefEq.lean}).
\end{proof}

\paragraph{Contrast with definitional proof irrelevance.}
Even without recursive predicates, \citet{abel-coquand2020} proved that combining \tttrule{Def-PI} with impredicative $\Prop$ and large $\Id$-elimination yields a non-normalizing closed term by equating an open equality proof with $\refl$. With $\tttAcc$-recursion, \tttrule{Def-PI} breaks normalization and transitivity even without impredicativity:

\begin{proposition}[Failure of normalization and transitivity for $\tttAcc$ under \tttrule{Def-PI}~\cite{carneiro2019,gilbert2019}]\label{prop:def-pi-failure}
Extend the declarative theory with Rule~\tttrule{Def-PI} ($p \defeq q$ for $\ctx \vdash p, q : P$ and $\ctx \vdash P : \Prop$).
\begin{enumerate}
\item \emph{Loss of strong normalization:} Let $A := \Nat \to \Setu$, $R := \tttlam{y}{A}\allowbreak\tttlam{x}{A} \Id\,A\,x\,(\tttlam{n}{\Nat} y\,(n+1) \to \tttFalse)$, and define $B\,(x : A)\,(p : \tttAcc\,A\,R\,x) : \Setu$ by $\rec_{\tttAcc}$ with minor premise
\[
\begin{aligned}
  &\tttlam{x}{A}\tttlam{h}{(\Pi y{:}A.\,R\,y\,x \to \tttAcc\,A\,R\,y)} \\[-2pt]
  &\quad\tttlam{\mathit{ih}}{(\Pi y{:}A.\,\Pi r{:}R\,y\,x.\,B\,y\,(h\,y\,r))} \\[-2pt]
  &\qquad\bigl(x\,0 \to \mathit{ih}\,(\tttlam{n}{\Nat} x\,(n+1))\,(\refl\,x)\,0\bigr).
\end{aligned}
\]
In the context $p_0 : \tttAcc\,A\,R\,(\tttlam{\_}{\Nat}\tttFalse)$, set $a_0 := \tttlam{\_}{\Nat} B\,\allowbreak(\tttlam{\_}{\Nat}\tttFalse)\,p_0 : A$ and $h_0 := \tttlam{y}{A}\allowbreak\tttlam{\_}{R\,y\,a_0} p_0$. Because $R\,a_0\,a_0 = \Id\,A\,a_0\,\allowbreak(\tttlam{n}{\Nat}\allowbreak B\,(\tttlam{\_}{\Nat}\tttFalse)\,p_0 \to \tttFalse)$, unfolding $B$ on $\mathsf{intro}\,a_0\,h_0$ uses \tttrule{Def-PI} at $\tttAcc\,A\,R\,a_0$ to identify $p_0 \defeq \mathsf{intro}\,a_0\,h_0$, making $\omega := \tttlam{z}{B\,a_0\,p_0} z\,z\,(\refl\,a_0)$ well-typed and $\Omega := \omega\,\omega$ diverge.
\item \emph{Non-transitivity of algorithmic conversion:} Any syntax-directed conversion test that reduces $\rec_{\tttAcc}$ on constructor terms while equating arbitrary terms of type $\tttAcc\,A\,R\,x$ fails transitivity on open proofs $p : \tttAcc\,A\,R\,x$, because $\rec_{\tttAcc}\,A\,R\,C\,f\,x\,p$ and $\rec_{\tttAcc}\,A\,R\,C\,f\,x\,\allowbreak(\mathsf{intro}\,x\,h)$ are both equated via $p \defeq \mathsf{intro}\,x\,h$, whereas the right-hand side unfolds once more to $f\,x\,h\,(\cdots)$ which cannot be joined with the neutral left-hand side without synthesizing a proof of $R\,y\,x$.
\end{enumerate}
\end{proposition}

In Eigenius, untyped joinability (Definition~\ref{def:defeq}) blocks both the Abel--Coquand and $\tttAcc$ counterexamples: $\Jelim$ on a variable or $\irrel$ is neutral ($e \not\defeq \refl\,a$), and $p_0 \not\defeq \mathsf{intro}\,a_0\,h_0$.

\subsection{Witnessed Proof Irrelevance}
\label{sec:kernel-pi}

Rather than placing proof irrelevance in conversion, Eigenius declares a single global axiom $\mathsf{axiom}\ \irrel : \Pi P{:}\Prop.\allowbreak\,\Pi p{:}P.\allowbreak\,\Pi q{:}P.\allowbreak\,\Id\,P\,p\,q$ in $\tttEnv$ (whose type is itself a closed proposition in $\Prop$ by Rule~\tttrule{$\Pi$-F}). Instantiating $\irrel$ via \tttrule{Const} and three applications of \tttrule{$\Pi$-E} yields the derived proof-irrelevance rule:
\[
  \inferrule*[right=\tttrule{PI}]
    {\ctx \vdash P : \Prop \\ \ctx \vdash p : P \\ \ctx \vdash q : P}
    {\ctx \vdash \irrel\,P\,p\,q : \Id\,P\,p\,q}
\]
This design achieves three simultaneous properties required by a verified scientific store:
\begin{enumerate}
\item \textbf{Retained distinctness of proof witnesses.} For any two proofs $p, q : P$ of a proposition $P : \Prop$ (such as $p_{\mathrm{CRISPR}}$ and $p_{\mathrm{RNAi}}$ in \S\ref{sec:intro-wrn}), $p \not\defeq q$ unless $p$ and $q$ reduce to the same normal form. The store and its institutional comorphisms (\S\ref{sec:inst-env}) can therefore inspect $p$ and $q$ as distinct syntax trees.
\item \textbf{Internalized propositional equality of proofs.} Whenever a dependent data type (e.g., a subset type $\Sigma x{:}A.\,P\,x$) or theorem requires transporting a motive $C : P \to \Srt\,\ell'$ from $p$ to $q$, the term
\[
  \mathsf{tr}_C(\irrel\,P\,p\,q, d) := \Jelim\,p\,(\tttlam{y}{P}\tttlam{\_}{\Id\,P\,p\,y} C\,y)\,d\,q\,(\irrel\,P\,p\,q) : C\,q
\]
is well-typed in the kernel. Because Rule~\tttrule{Id-$\iota$} fires only when the identity proof is of the form $\refl\,a'$, and $\irrel$ is an axiom with no $\delta$-rule, this transport term is a \emph{neutral normal form} that explicitly records the use of proof irrelevance between $p$ and $q$.
\item \textbf{Hyperintensionality of propositions.} In Rule~\tttrule{PI}, the premise $P : \Prop$ types the \emph{proofs} $p, q : P$, not $P$ itself~\cite{pfenning2001irrelevance}: propositions $P, Q : \Prop$ inhabit $\Prop : \Setu$ ($\Setu \neq \Prop$). Thus $\irrel$ never inhabits $\Id\,\Prop\,P\,Q$, and $\mathsf{And}\,P\,Q \defeq \mathsf{And}\,P'\,Q'$ holds only when $P \defeq P'$ and $Q \defeq Q'$.
\end{enumerate}

\begin{remark}[Why $\irrel$ cannot be replaced by \tttrule{PI-Refl}]\label{rem:pi-refl-fails}
Replacing the axiom $\irrel$ by a modified reflexivity rule \tttrule{PI-Refl} ($\ctx \vdash \refl\,p : \Id\,P\,p\,q$ for $\ctx \vdash P : \Prop$ and $\ctx \vdash p, q : P$) breaks subject reduction under untyped \tttrule{Id-$\iota$}: for $p \not\defeq q : P$ and $C : \Pi y{:}P.\,\Id\,P\,p\,y \to \Setu$ with $d : C\,p\,(\refl\,p)$, $\Jelim\,p\,C\,d\,q\,(\refl\,p)$ has type $C\,q\,(\refl\,p)$, whereas its \tttrule{Id-$\iota$}-reduct $d$ has type $C\,p\,(\refl\,p) \not\defeq C\,q\,(\refl\,p)$.
\end{remark}

\section{Subsumption \& Bidirectional Typing}
\label{sec:subsumption}

Because Eigenius types both mathematical constructions across cumulative universes and open-world knowledge-graph resources across ontology class hierarchies (\S\ref{sec:inst-kg}), subtyping enters the kernel through a single unified subsumption judgment $\ctx \vdash A \subty B$.

\subsection{Unified Subsumption Judgment}
\label{sec:subsumption-unified}

Figure~\ref{fig:subsumption} presents the declarative rules for $\ctx \vdash A \subty B$. It combines four sources of subtyping:
(1)~definitional equality (\tttrule{S-Conv}) and transitivity (\tttrule{S-Trans});
(2)~universe cumulativity (\tttrule{S-Cum}) governed by $m \cumle n$, so that $\Setu \subty \Typeu_1 \subty \cdots$ while $\Prop$ neither lies below nor above any data universe ($\Prop \not\subty \Setu$; were $\Prop \subty \Setu$ permitted, substituting $\Phi : \Prop$ for $X : \Setu$ in the data function $\tttlam{x}{X} x$ would turn it into a proof $\tttlam{x}{\Phi} x$, destroying Sort Agreement in \S\ref{sec:semantics});
(3)~covariant codomain subsumption on dependent function types with definitionally equal domains (\tttrule{S-$\Pi$}); and
(4)~declared ontology class inclusions (\tttrule{Class-Sub} along the reflexive--transitive subclass preorder $\kganc$ and \tttrule{Class-Ent} into $\Entity$) together with width/depth record subtyping (\tttrule{Rec-Sub}, detailed in \S\ref{sec:inst-kg}).

\begin{figure}[t]
\footnotesize
\begin{mathpar}
\inferrule*[right=\tttrule{S-Conv}]
  {A \defeq B}
  {\ctx \vdash A \subty B}
\and
\inferrule*[right=\tttrule{S-Trans}]
  {\ctx \vdash A \subty B \\ \ctx \vdash B \subty C}
  {\ctx \vdash A \subty C}
\and
\inferrule*[right=\tttrule{S-Cum}]
  {\ctx \tttctx \\ m \cumle n}
  {\ctx \vdash \Srt\,m \subty \Srt\,n}
\\
\inferrule*[right=\tttrule{S-$\Pi$}]
  {\ctx \vdash A_1 : \Srt\,\ell_1 \\ \ctx \vdash A_2 : \Srt\,\ell_2 \\ A_1 \defeq A_2 \\
   \ctx, x{:}A_2 \vdash B_1 \subty B_2}
  {\ctx \vdash \Pi x{:}A_1.\,B_1 \subty \Pi x{:}A_2.\,B_2}
\\
\inferrule*[right=\tttrule{Class-Sub}]
  {\ctx \tttctx \\ C \kganc D}
  {\ctx \vdash C \subty D}
\and
\inferrule*[right=\tttrule{Class-Ent}]
  {\ctx \tttctx \\ C \in \mathcal{C}_{\mathcal{G}}}
  {\ctx \vdash C \subty \Entity}
\and
\inferrule*[right=\tttrule{Rec-Sub}]
  {\kglabels(R') \subseteq \kglabels(R) \\ \ctx, \kgtel(R) \vdash A_\ell \subty A'_\ell\ (\forall \ell \in \kglabels(R')) \\ w \le w'}
  {\ctx \vdash R \subty R'}
\end{mathpar}
\vspace{-8pt}
\caption{The unified subsumption judgment $\ctx \vdash A \subty B$, combining conversion, universe cumulativity, $\Pi$-codomain covariance, ontology class hierarchies, and record width/depth subtyping (where $w, w' \ge 1$ are the implicit level bounds of $R, R'$, \S\ref{sec:inst-kg}).}
\Description{The unified subsumption judgment combining conversion, universe cumulativity, dependent product codomain covariance, ontology class hierarchies, and record width/depth subtyping.}
\label{fig:subsumption}
\vspace{-6pt}
\end{figure}

\paragraph{Why convertible domains in \tttrule{S-$\Pi$} and explicit $\eta$ for contravariance.}
Following Luo's ECC~\cite{luo1994} and Coq, Rule~\tttrule{S-$\Pi$} requires the domains to be definitionally equal ($A_1 \defeq A_2$) rather than contravariant ($A_2 \subty A_1$). This choice has two decisive metatheoretic advantages:
\begin{enumerate}
\item \emph{Direct set-theoretic inclusion without coercions (\S\ref{sec:semantics}).} In our set-theoretic model, a data abstraction $\tttlam{x}{A_1} b$ denotes the set-theoretic graph of its function on $\llbracket A_1 \rrbracket$. When $A_1 \defeq A_2$, $\llbracket A_1 \rrbracket = \llbracket A_2 \rrbracket$, so any graph on $\llbracket A_1 \rrbracket$ taking values in $\llbracket B_1(x) \rrbracket \subseteq \llbracket B_2(x) \rrbracket$ is literally an element of $\llbracket \Pi x{:}A_2.\,B_2 \rrbracket$. If $A_2 \subty A_1$ were a strict subset, a function graph on $\llbracket A_1 \rrbracket$ would have to be restricted to $\llbracket A_2 \rrbracket$, requiring a semantic coercion.
\item \emph{Direct structural transitivity elimination.} In a dependent calculus with contravariant domains, composing $\Pi x{:}A_1.\,B_1 \subty \Pi x{:}A_2.\,B_2$ and $\Pi x{:}A_2.\,B_2 \subty \Pi x{:}A_3.\,B_3$ requires \emph{context narrowing} from $x{:}A_2$ to $x{:}A_3$ in the derivation of $B_1 \subty B_2$. Because narrowing a dependent derivation inserts subsumption steps at every variable occurrence, neither derivation height nor $\Pi$-nesting depth is preserved, forcing transitivity elimination to rely on strong normalization~\cite{aspinall-compagnoni2001}. With $A_1 \defeq A_2 \defeq A_3$, transitivity is proved by a direct structural induction (Lemma~\ref{lem:k1-trans}) without normalization.
\end{enumerate}
At the same time, contravariant domain adaptation is not lost: whenever $\ctx \vdash A_2 \subty A_1$, any function $f : \Pi x{:}A_1.\,B_1$ adapts to the narrower domain $A_2$ by explicit $\eta$-expansion:

\begin{proposition}[Contravariant domain adaptation via $\eta$-expansion]\label{prop:eta-contra}
If $\ctx \vdash f : \Pi x{:}A_1.\,B_1$, $\ctx \vdash A_2 : \Srt\,\ell_2$, $\ctx \vdash A_2 \subty A_1$, and $\ctx, x{:}A_2 \vdash B_1 \subty B_2 : \Srt\,\ell'_2$, then $\ctx \vdash \tttlam{x}{A_2} f\,x : \Pi x{:}A_2.\,B_2$.
\end{proposition}
\begin{proof}
In $\ctx, x{:}A_2$, \tttrule{Sub} gives $x : A_1$; \tttrule{$\Pi$-E} yields $f\,x : B_1$, hence $f\,x : B_2$ by \tttrule{Sub}, and \tttrule{$\Pi$-I} gives $\ctx \vdash \tttlam{x}{A_2} f\,x : \Pi x{:}A_2.\,B_2$.
\end{proof}

Notice also that $\Sigma$-types have no structural subsumption rule in Figure~\ref{fig:subsumption}: in $\Sigma X{:}\Srt\,\ell_1.\,B(X)$, $B(X)$ may compute on $X$ by large elimination or quantification in $\Setu$, which is non-monotone in $\ell_1$.

\subsection{Bidirectional Kernel Fragments (\texorpdfstring{$\mathcal{K}_0$ and $\mathcal{K}_1$}{K0 and K1})}
\label{sec:subsumption-bidir}

Declarative typing (Figure~\ref{fig:kernel-typing}) is not syntax-directed because \tttrule{Conv} and \tttrule{Sub} can apply at any node and \tttrule{S-Trans} guesses an intermediate type $B$. The Eigenius kernel implements a syntax-directed bidirectional split into a \emph{type-synthesis} judgment $\ctx \vdash a \infr A$ and a \emph{type-checking} judgment $\ctx \vdash a \chk A$ (Figure~\ref{fig:bidir}).

\begin{figure}[t]
\footnotesize
\begin{mathpar}
\inferrule*[right=\tttrule{Inf-Var}]
  {(x{:}A) \in \ctx}
  {\ctx \vdash x \infr A}
\and
\inferrule*[right=\tttrule{Inf-Sort}]
  { }
  {\ctx \vdash \Srt\,\ell \infr \Srt\,(\lsucc\,\ell)}
\and
\inferrule*[right=\tttrule{Inf-Ann}]
  {\ctx \vdash A \infr_{\mathsf{s}} \ell \\ \ctx \vdash a \chk A}
  {\ctx \vdash \tttann{a}{A} \infr A}
\\
\inferrule*[right=\tttrule{Inf-$\Pi$}]
  {\ctx \vdash A \infr_{\mathsf{s}} \ell_1 \\ \ctx, x{:}A \vdash B \infr_{\mathsf{s}} \ell_2}
  {\ctx \vdash \Pi x{:}A.\,B \infr \Srt\,(\limax\,\ell_1\,\ell_2)}
\and
\inferrule*[right=\tttrule{Inf-Lam}]
  {\ctx \vdash A \infr_{\mathsf{s}} \ell_1 \\ \ctx, x{:}A \vdash b \infr B}
  {\ctx \vdash \tttlam{x}{A} b \infr \Pi x{:}A.\,B}
\\
\inferrule*[right=\tttrule{Chk-Lam}]
  {C \defeq_K \Pi x{:}A'.\,B \\ \ctx \vdash A \infr_{\mathsf{s}} \ell \\ A \defeq_K A' \\ \ctx, x{:}A \vdash b \chk B}
  {\ctx \vdash \tttlam{x}{A} b \chk C}
\and
\inferrule*[right=\tttrule{Inf-App}]
  {\ctx \vdash f \infr C \\ C \defeq_K \Pi x{:}A.\,B \\ \ctx \vdash a \chk A}
  {\ctx \vdash f\,a \infr B[\tttann{a}{A}/x]}
\\
\inferrule*[right=\tttrule{Chk-Pair}]
  {C \defeq_K \Sigma x{:}A.\,B \\ \ctx \vdash a \chk A \\ \ctx \vdash b \chk B[\tttann{a}{A}/x]}
  {\ctx \vdash \tttpair{a}{b} \chk C}
\and
\inferrule*[right=\tttrule{Inf-Fst}]
  {\ctx \vdash p \infr C \\ C \defeq_K \Sigma x{:}A.\,B}
  {\ctx \vdash \fst\,p \infr A}
\and
\inferrule*[right=\tttrule{Inf-Snd}]
  {\ctx \vdash p \infr C \\ C \defeq_K \Sigma x{:}A.\,B}
  {\ctx \vdash \snd\,p \infr B[\fst\,p/x]}
\\
\inferrule*[right=\tttrule{Chk-Refl}]
  {C \defeq_K \Id\,A\,x\,y \\ \ctx \vdash a \chk A \\ x \defeq_K a \defeq_K y}
  {\ctx \vdash \refl\,a \chk C}
\and
\inferrule*[right=\tttrule{Chk-Sub}]
  {\ctx \vdash a \infr A \\ \ctx \vdash A \subty_K B}
  {\ctx \vdash a \chk B}
\end{mathpar}
\vspace{-8pt}
\caption{Bidirectional synthesis ($\ctx \vdash a \infr A$) and checking ($\ctx \vdash a \chk C$) rules for $\mathcal{K}_0$ and $\mathcal{K}_1$, writing $\ctx \vdash A \infr_{\mathsf{s}} \ell$ for $\ctx \vdash A \infr S$ with $S \defeq_K \Srt\,\ell$ (expected types and named reducts well-formed; constants, $\Sigma$/$\Id$, $\Jelim$ omitted).}
\Description{Bidirectional synthesis and checking rules for the kernel fragments K0 and K1.}
\label{fig:bidir}
\vspace{-6pt}
\end{figure}

We study two nested bidirectional fragments isolating the kernel's metatheoretic obligations:
\begin{itemize}
\item \textbf{Fragment $\mathcal{K}_0$:} Subsumption $A \subty_K B$ at \tttrule{Chk-Sub} holds iff $A \defeq_K B$, or $A \defeq_K \Srt\,m$ and $B \defeq_K \Srt\,n$ with $m \cumle n$.
\item \textbf{Fragment $\mathcal{K}_1$:} Extends $\subty_K$ with $\Pi$-subsumption ($A \defeq_K \Pi x{:}A_1.\,B_1$, $B \defeq_K \Pi x{:}A_2.\,B_2$, $A_1 \defeq_K A_2$, $\ctx, x{:}A_2 \vdash B_1 \subty_K B_2$) and the class, singleton, and join rules of Figures~\ref{fig:subsumption} and~\ref{fig:kg-rules} (reading types through reducts).
\end{itemize}
Notice that in \tttrule{Inf-App} and \tttrule{Chk-Pair}, substituting a checked argument $a \chk A$ into a dependent codomain $B$ wraps $a$ in an explicit annotation $B[\tttann{a}{A}/x]$: $a$ may be a check-only constructor such as $\tttpair{u}{v}$, whereas $x$ inside $B$ may stand at an inference position such as $\fst\,x$, and \tttrule{Ann-$\tttstep$} keeps $\tttann{a}{A}$ invisible to $\defeq$.

\begin{lemma}[Transitivity of algorithmic subsumption]\label{lem:k1-trans}
In both $\mathcal{K}_0$ and $\mathcal{K}_1$, if $\ctx \vdash A \subty_K B$ and $\ctx \vdash B \subty_K C$, then $\ctx \vdash A \subty_K C$; consequently, if $\ctx \vdash a \chk A$ and $\ctx \vdash A \subty_K B$ for a well-formed type $B$, then $\ctx \vdash a \chk B$.
\end{lemma}
\begin{proof}
By structural induction on $\ctx \vdash A \subty_K B$ and inner induction on $\ctx \vdash B \subty_K C$ (handling \tttrule{Join-I} against \tttrule{Join-L}, while graph-type bounds isolate joins and singletons from $\Srt, \Pi, \Sigma$, \code{SubK.data\_right}). By Theorem~\ref{thm:confluence}(4), cumulativity meets only joinability or cumulativity ($\cumle$ transitive), never $\Pi$. When two $\Pi$-steps meet at $B$, $\Pi$-injectivity (Theorem~\ref{thm:confluence}(3)) gives $A_1 \defeq_K A_2 \defeq_K A_3$; auxiliary context conversion (\code{Bidir.ctxConv}: replacing $x{:}A_2$ by $x{:}A_3 \defeq_K A_2$ preserves $\chk, \subty_K$ and preserves $\infr$ up to $\defeq_K$) transfers the first premise to $x{:}A_3$, where the induction hypothesis composes the codomains. The second claim inducts on $\ctx \vdash a \chk A$.
\end{proof}

\paragraph{Annotated reduction and subject reduction.}
In a bidirectional system, plain reduction $\tttstep$ can destroy typability if it strips a type annotation from an application head---for instance, $\tttann{\tttlam{x}{A}\tttpair{x}{x}}{A \to \Sigma y{:}A.\,A}\,a$ synthesizes $\Sigma y{:}A.\,A$, whereas its plain reduct $\tttpair{a}{a}$ cannot synthesize any type. We therefore define \emph{annotated reduction} $\tttstep_a$ (where $V \defeq_K \Pi x{:}A'.\,B$ in the application rules and $V \defeq_K \Sigma x{:}A.\,B$ in the projection rules), which keeps every term landing in an inference position annotated:
\begin{align*}
  (\tttlam{x}{A} b)\,a &\;\tttstep_a\; b[\tttann{a}{A}/x], \\
  \tttann{\tttlam{x}{A} b}{V}\,a &\;\tttstep_a\; \tttann{b[\tttann{a}{A}/x]}{B[\tttann{a}{A}/x]}, \\
  \tttann{u}{V}\,a &\;\tttstep_a\; \tttann{u\,\tttann{a}{A'}}{B[\tttann{a}{A'}/x]} \quad (\text{in }\mathcal{K}_1), \\
  \fst\,\tttann{\tttpair{a}{b}}{V} &\;\tttstep_a\; \tttann{a}{A}, \;
  \snd\,\tttann{\tttpair{a}{b}}{V} \tttstep_a \tttann{b}{B[\tttann{a}{A}/x]},
\end{align*}
together with $\delta$-unfolding to $\tttann{t[\bar\ell/\bar u]}{A[\bar\ell/\bar u]}$, annotated $\iota$- and $\Jelim$-reduction, and erasure $\tttann{a}{A} \tttstep_a a$ when $\tttann{a}{A}$ already infers or stands in checking position.

\begin{theorem}[Bidirectional subject reduction, erasure soundness, and annotation completeness]\label{thm:bidir-metatheory}
Over any well-formed $\tttEnv$ in $\mathcal{K}_1$:
\begin{enumerate}
\item \emph{Subject reduction for $\tttstep_a$:} If $\ctx \vdash t \infr T$ and $t \tttstep_a t'$, then $\ctx \vdash t' \infr T'$ with $T \defeq_K T'$; if $\ctx \vdash t \chk T$ and $t \tttstep_a t'$, then $\ctx \vdash t' \chk T$.
\item \emph{Reduction lifting:} For typable $t$, $|t| \tttstep^* s$ iff $t \tttstep_a^* t'$ for some $t'$ with $|t'| = s$.
\item \emph{Soundness and completeness:} For $\ctx \tttctx$, $\ctx \vdash a \infr A$ or $\ctx \vdash a \chk A$ in $\mathcal{K}_1$ implies $|\ctx| \vdash |a| : |A|$ declaratively (and $\ctx \vdash a : A$ since $\tttann{u}{U} \defeq u$). Conversely, $\ctx \vdash a : A$ yields $a', A', \ctx'$ with $|a'| = |a|$, $|A'| = |A|$, $a \defeq a'$, $A \defeq A'$, and $\ctx' \vdash a' \chk A'$ in $\mathcal{K}_1$.
\end{enumerate}
\end{theorem}
\begin{proof}[Proof sketch]
(1)~Inducts on the derivation via Lemma~\ref{lem:k1-trans}; in $\mathcal{K}_1$, $\tttann{u}{V}\,a \tttstep_a \tttann{u\,\tttann{a}{A'}}{B[\tttann{a}{A'}/x]}$ handles heads whose inferred type $S \subty_K V$ has a smaller codomain than $V$. (2)~Pushing annotations outward along spines exposes the $\beta/\Sigma/\iota$-redex while preserving $|t|$. (3)~Soundness erases annotations; completeness inducts on $\ctx \vdash a : A$, using Lemma~\ref{lem:k1-trans} and annotating heads $\tttann{f'}{\Pi x{:}A'.\,B'}\,a'$ (\code{AnnotatedReduction.lean}, \code{Completeness.lean}).
\end{proof}

\subsection{Algorithmic Checking \& Comparison with Coquand's \emph{Unit}}
\label{sec:subsumption-unit}

In Figure~\ref{fig:bidir}, premises such as $C \defeq_K \Pi x{:}A.\,B$ still quantify existentially over joinable reducts. Following \citet{coquand1996}, our algorithmic systems $\mathcal{A}_0$ and $\mathcal{A}_1$ (\code{Checking/\allowbreak Algorithmic.lean}, \code{Procedure.lean}) compile $\mathcal{K}_0$ and $\mathcal{K}_1$ into an executable CPS checker (\code{infer}/\code{check}/\code{sub}) that tries constructor-directed checking (\code{Query.chkForm}) before falling back via \code{orElse} to~\tttrule{Chk-Sub}. Every type premise $C \defeq_K \Pi x{:}A.\,B$ (and $\Srt, \Sigma, \Id$) is decided by deterministic weak-head reduction $|C| \chkred \Pi x{:}A.\,B$, justified by confluence and WHNF stability (\code{Rigid.join}). Conversion $a \defeq_K b$ consults no types: it is decided either by structural WHNF recursion (\code{ConvRun}) or by untyped \code{NbE}~\cite{aehlig-joachimski2004}, avoiding the type-directed $\eta$-readback of typed NbE~\cite{nbe}. Soundness and completeness relative to termination hold unconditionally (\code{Alg.sound}, \code{ConvRun.correct}, \code{Run.correct})---so Grothendieck consistency (\S\ref{sec:semantics}) needs no normalization hypothesis---and both \code{ConvRun} and \code{NbE} terminate on all strongly normalizing erasures (\code{ConvRun.total}, \code{Nbe.total} in \code{Checking/\allowbreak Termination.lean}).

\begin{conjecture}[Strong normalization and canonicity boundary]\label{conj:sn}
Over any admissible $\tttEnv$, every declaratively typable $\ctx \vdash a : A$ has strongly normalizing erasure $|a|$ under $\tttstep$ (a Girard--Werner obligation at the strength of cumulative CIC + $\tttAcc$~\cite{barras2010,werner1997}, left open here), making $\mathcal{A}_0, \mathcal{A}_1$ total decision procedures. Closed data terms eliminating the axiom $\irrel$ via $\Jelim$ into $\Setu$ remain neutral transports (\S\ref{sec:kernel-pi}), whereas the $\irrel$-free data fragment ($\Evid_{\jlcs}, \supp, \jlsw, \kgcanon$ in \S\ref{sec:institutions}) computes by $\beta\delta\iota$.
\end{conjecture}

\begin{table}[t]
\centering\footnotesize
\setlength{\tabcolsep}{3pt}
\caption{Architectural comparison between Coquand's \emph{Unit}~\cite{coquand2026unit} and the Eigenius kernel checker ($\mathcal{K}_1 / \mathcal{A}_1$).}
\label{tab:unit-comparison}
\begin{tabularx}{\columnwidth}{@{}l >{\raggedright\arraybackslash}X >{\raggedright\arraybackslash}X@{}}
\toprule
\textbf{Aspect} & \textbf{Coquand's \emph{Unit}~\cite{coquand2026unit}} & \textbf{Eigenius Kernel ($\mathcal{K}_1 / \mathcal{A}_1$)} \\
\midrule
TCB \& data & Explicit bidir.\ terms, recursors; no unification or tactics & Explicit bidir.\ certificates, $\rec_D$; no unification or tactics \\
$\delta$-unfolding & Compare shared head args first; unfold newer \code{def} (no heuristics) & Confluence-invariant transparent \code{def}s; sealed opaque \code{def}s \\
$\eta$-equality & Untyped WHNF $\eta$ on unannotated $\lambda$ and pairs & Omitted from $\defeq$ (annotated $\lambda$, domain graphs, untyped \code{NbE}) \\
Universes & Cumulative predicative $\mathcal{U}_\ell$ (level exprs.) & Isolated impredicative $\Prop$ + $\irrel$; cumulative $\Setu \subty \Typeu_k$ \\
Subsumption & Universe cumulativity & Cumulativity, $\Pi$-codomains, ontology classes \& TTR records \\
Modularity & Automath-style parameterized declaration segments & Signature morphisms $\sigma$ + cross-logic comorphisms (\S\ref{sec:inst-env}) \\
\bottomrule
\end{tabularx}
\end{table}

Table~\ref{tab:unit-comparison} situates this design alongside Coquand's \emph{Mini-TT}~\cite{minitt} and \emph{Unit}~\cite{coquand2026unit}. Eigenius adopts \emph{Unit}'s core thesis: rather than embedding elaboration heuristics, metavariables, pattern-matching compilation, or type-directed \tttrule{Def-PI} inside the TCB, the kernel checks explicit bidirectional certificates and recursors deterministically. By confluence and WHNF stability, \emph{Unit}'s uniform definition rule---comparing shared defined heads by arguments first and unfolding the newer definition on mismatch---decides the exact same relation $\defeq_K$ as eager WHNF unfolding, while Eigenius adds \emph{opaque} definitions to seal domain abstractions. Each remaining departure serves scientific epistemics:
(i)~whereas \emph{Unit} tests $\eta$ untyped on unannotated abstractions~\cite{coquand1996}, Eigenius carries domain annotations $\tttlam{x}{A} b$, where untyped $\eta$ breaks $\beta\eta$-confluence ($\tttlam{x}{A}(\tttlam{y}{B}y)\,x$) and equates $\tttlam{x}{A_2} f\,x$ with $f : \Pi x{:}A_1.\,B_1$ across distinct semantic domains $\llbracket A_2\rrbracket \subsetneq \llbracket A_1\rrbracket$ (Proposition~\ref{prop:eta-contra});
(ii)~isolating impredicative $\Prop$ ($\Prop \not\subty \Setu$) reconciles impredicative invariants and witnessed $\irrel$ with distinct proof terms ($p_{\mathrm{CRISPR}} \not\defeq p_{\mathrm{RNAi}}$); and
(iii)~extending~\tttrule{Chk-Sub} and segment morphisms to ontology classes, canonical TTR records, and institutional comorphisms (\S\ref{sec:institutions}) types open-world multi-verifier stores without coercions.

\section{Constructive Institutions, Justification \& Knowledge Graphs}
\label{sec:institutions}

No single deductive calculus covers every mode of reasoning in an empirical study: statistical inference plans, external theorem provers (Lean~4, SMT solvers), justification-logic warrant evaluators, and open-world knowledge-graph schemas operate over distinct signatures and witness structures. Eigenius unifies them by organizing environment theories into a three-level institutional hierarchy (\S\ref{sec:inst-env}), internalizing an empirical justification logic over kernel propositions (\S\ref{sec:inst-jl}), and embedding a dependent record calculus for typed knowledge graphs (\S\ref{sec:inst-kg}).

\subsection{Environment Theories as Constructive Institutions}
\label{sec:inst-env}

Recall that a Goguen--Burstall \emph{institution}~\cite{goguen1992} $\instI = (\Sign,\allowbreak \Sen,\allowbreak \Mod,\allowbreak \models)$ equips a category of signatures with a covariant sentence functor $\Sen : \Sign \to \SetC$, a contravariant model functor $\Mod : \Sign^{\mathrm{op}} \to \Cat$, and satisfaction relations ${\models_\Sigma} \subseteq |\Mod(\Sigma)| \times \Sen(\Sigma)$ satisfying $M' \models_{\Sigma'} \Sen(f)(\varphi) \allowbreak\iff\allowbreak \Mod(f)(M') \models_\Sigma \varphi$. Because an institution or Tarskian consequence relation records only \emph{that} a sentence holds and not \emph{by which witness}, Eigenius stratifies its multi-logic architecture across three levels related by forgetful functors:
\[
  \instPC \;\xrightarrow{\;\mathit{ent}\;}\; \Ent \;\xrightarrow{\;\instU\;}\; \piIns^{\instco} \;\underset{\instF}{\overset{\instG}{\rightleftarrows}}\; \Ins^{\instco}.
\]
At the primary \emph{witness level} ($\instPC$), each hosted logic is a \emph{proof calculus} in the sense of Meseguer~\cite[Definition~12]{meseguer1989}: a tuple $\instP = (\Sign, \Sen, \vdash, P, \instPr, \pi)$ where $(\Sign, \Sen, \vdash)$ is a compact entailment system, $P : \Thz \to \mathbf{Struct}_{\instP}$ assigns to each theory $T = (\Sigma, \Gamma)$ its proof-theoretic structure (a multicategory of derivations), $\proofs = \instPr \circ P : \Thz \to \SetC$ extracts the set of witnesses, and $\pi : \proofs \Rightarrow \Sen$ is a natural projection onto the theorem closure $\instclo{\Gamma} = \{\varphi \mid \Gamma \vdash_\Sigma \varphi\}$.

\paragraph{The Eigenius kernel as a proof calculus $\instK$.}
A \emph{kernel signature} $\Sigma \in |\Sign_{\instK}|$ is any well-formed global environment (\code{KSign} in \code{Institutions/\allowbreak Kernel.lean}, including foundational declarations such as $\irrel$ and defined lemmas, while theory-specific empirical hypotheses are placed in $\Gamma$). A \emph{signature morphism} $\sigma : \Sigma \to \Sigma'$ assigns to each constant $\tttinst{c}{\bar u} : A$ of $\Sigma$ a closed term $\sigma(c)$ of $\Sigma'$ such that its homomorphic extension $\insthat{\sigma}$ preserves typing ($\Sigma'; \cdot \vdash \sigma(c) : \insthat{\sigma}(A)$), $\delta$-unfolding ($\sigma(c) \defeq \insthat{\sigma}(b)$), and $\iota$-computation rules up to $\defeq$ in $\Sigma'$. Store updates (extending an environment with new declarations, renaming constants, or instantiating parameters) are signature morphisms in $\Sign_{\instK}$.

The sentence set $\Sen_{\instK}(\Sigma)$ consists of closed propositions ($\Sigma; \cdot \vdash \varphi : \Prop$). For a theory $T = (\Sigma, \Gamma)$ with $\Gamma \subseteq \Sen_{\instK}(\Sigma)$, each axiom $\varphi \in \Gamma$ is assigned a canonical hypothesis variable $\instax{\varphi} : \varphi$. A \emph{kernel witness} in $\proofs_{\instK}(T)$ is a pair $(t, \varphi)$ where $\Sigma; \instax{\Gamma_0} \vdash t : \varphi$ for some finite $\Gamma_0 \subseteq \Gamma$, taken up to $\alpha$-equivalence of bound variables and \emph{not} up to $\defeq$ or propositional identity $\Id$, with $\pi_T(t, \varphi) = \varphi$.

\begin{theorem}[The kernel proof calculus and the proof-relevance boundary]\label{thm:kernel-pc}
\leavevmode
\begin{enumerate}
\item \emph{Proof calculus:} $\instK = (\Sign_{\instK}, \Sen_{\instK}, \vdash, P_{\instK}, \instPr, \pi)$ is a compact proof calculus whose entailment is Tarskian and preserved along signature morphisms ($\Gamma \vdash_\Sigma \varphi \implies \insthat{\sigma}(\Gamma) \vdash_{\Sigma'} \insthat{\sigma}(\varphi)$). Moreover, cut of $\Sigma; \instax{\Delta}, \instax{\varphi_1}{:}\varphi_1, \ldots, \instax{\varphi_k}{:}\varphi_k \vdash s : \psi$ with $\Sigma; \instax{\Delta} \vdash t_j : \varphi_j$ is realized by $(\tttlam{\instax{\varphi_1}}{\varphi_1}\cdots\tttlam{\instax{\varphi_k}}{\varphi_k} s)\,t_1\cdots t_k$, requiring no conversion checks beyond syntactic identity of each $\varphi_j$.
\item \emph{Collapse under $\irrel$-quotienting:} Let $\instK_{=}$ quotient $\proofs_{\instK}(T)$ (over signatures with $\irrel$) by provable identity: $(t, \varphi) \sim (t', \varphi)$ iff $\Sigma; \instax{\Gamma_0} \vdash e : \Id\,\varphi\,t\,t'$ for some $e$ and finite $\Gamma_0 \subseteq \Gamma$. Then every $\pi_T$ is injective, and $\instK_{=} \cong \mathit{ent}(\instK)^\natural$ via the unit of $\mathit{ent} \dashv (\cdot)^\natural$, leaving at most one witness per theorem.
\end{enumerate}
\end{theorem}
\begin{proof}
(1)~By induction on derivations, $\insthat{\sigma}$ commutes with capture-avoiding substitution and maps $\beta/\Sigma/\Id/\delta/\iota$ steps to $\defeq$-equations in $\Sigma'$, which compose by transitivity of $\defeq$ (Theorem~\ref{thm:confluence}(2)). Multicategory composition in $P_{\instK}(T)$ is term substitution, which commutes with $\insthat{\sigma}$. For the cut redex, since $\psi$ and all $\varphi_j$ are closed, $\psi[t_j/\instax{\varphi_j}] = \psi$ syntactically. (2)~Given $(t, \varphi), (t', \varphi) \in \proofs_{\instK}(T)$ typed over $\instax{\Gamma_0}$ and $\instax{\Gamma_1}$, weakening to $\instax{\Gamma_0 \cup \Gamma_1}$ and applying Rule~\tttrule{PI} yields $\irrel\,\varphi\,t\,t' : \Id\,\varphi\,t\,t'$, so $(t, \varphi) \sim (t', \varphi)$ (\code{Institutions/Kernel.lean}).
\end{proof}

Theorem~\ref{thm:kernel-pc}(2) states the exact institutional meaning of our kernel design: keeping $\proofs_{\instK}(T)$ unquotiented retains distinct derivations ($p_{\mathrm{CRISPR}} \neq p_{\mathrm{RNAi}}$), whereas quotienting by $\irrel$ is an explicit forgetful projection to $\mathit{ent}(\instK)^\natural$ that mimics at the institutional level what \tttrule{Def-PI} would do silently inside conversion.

\paragraph{Importing external verdicts via comorphisms and axiom policies.}
External pipelines---such as the R statistical plans (\code{dd\_\allowbreak achilles\_\allowbreak limma}, \code{xenograft\_\allowbreak lme4}) in the WRN study~\cite{will-kasarskis2026-wrn}, SMT solvers, or a hosted Lean~4 checker $\mathbb{L}_A$---are modeled as proof calculi $\instP$ mapped into $\instK$ via \emph{maps of proof calculi}~\cite[Definition~33]{meseguer1989} $(\Phi, \alpha, \gamma) : \instP \to \instP'$, comprising an $\alpha$-sensible theory functor $\Phi : \Thz \to \Thz'$, sentence translation $\alpha : \Sen \Rightarrow \Sen' \circ \Phi$, and witness transport $\gamma : \proofs \Rightarrow \proofs' \circ \Phi$ obeying the \emph{transport invariant}:
\[
  \pi'_{\Phi T}(\gamma_T(p)) \;=\; \alpha_T(\pi_T(p)) \qquad \text{for all } T \in |\Thz| \text{ and } p \in \proofs(T).
\]
On an $\alpha$-simple extension $\Phi$, this yields provability preservation ($\Gamma \vdash_\Sigma \varphi \implies \alpha_\Sigma(\Gamma) \vdash'_{\Phi(\Sigma, \emptyset)} \alpha_\Sigma(\varphi)$) while keeping external R, Python, or SMT code \emph{outside the TCB}: a hallucinated step makes $\gamma_T(p)$ fail kernel checking at $\alpha_T(\pi_T(p))$.

When $p \in \proofs(T)$ depends on source axioms $\Gamma = \{\varphi_1, \allowbreak\ldots, \varphi_k\}$, three \emph{axiom policies} govern how they enter $\instK$ (Table~\ref{tab:axiom-policies}). Notably, \emph{Relativize} turns \S\ref{sec:intro-wrn}'s pseudoreplicated KM12 ANOVA into the axiom-free theorem $\alpha(\varphi_{\mathrm{indep}}) \to \Psi_{\mathrm{KM12}}$, preserving the conditional deduction even after replicate independence $\varphi_{\mathrm{indep}}$ is rejected:

\begin{table}[t]
\centering\small
\caption{The three institutional axiom policies for maps of proof calculi $(\Phi, \alpha, \gamma)$ into the Eigenius kernel $\instK$.}
\label{tab:axiom-policies}
\begin{tabularx}{\columnwidth}{@{}l l >{\raggedright\arraybackslash}X >{\raggedright\arraybackslash}p{1.65cm}@{}}
\toprule
\textbf{Policy} & \textbf{Cat.} & \textbf{Action of $\gamma_T$ on $\instax{\varphi}$} & \textbf{Warrant} \\
\midrule
\emph{Adopt} & $\Thz$ & Maps $\instax{\varphi}$ to target axiom $\instax{\alpha(\varphi)}$ in $\Phi(T) = (\Phi^\diamond\Sigma, \alpha(\Gamma))$ & Declared \\
\emph{Discharge} & $\Th$ & Substitutes a supplied target proof $t_{d_\chi}$ for each $\instax{\chi}$ ($\chi \in \alpha(\Gamma)$) & Inherited from $d_\chi$ \\
\emph{Relativize} & $\instThl$ & $\lambda$-abstracts hypotheses: $\lambda \vec h{:}\alpha(\vec\varphi).\allowbreak\,\gamma_T(t)[\vec h/\instax{\alpha(\vec\varphi)}]$ in $(\Phi^\diamond\Sigma, \emptyset)$ & Verified ($\alpha(\vec\varphi) \instimp \alpha(\psi)$) \\
\bottomrule
\end{tabularx}
\end{table}

\begin{theorem}[Soundness of Adopt, Discharge, and Relativize; Provability and Model Projections]\label{thm:policies-models}
\leavevmode
\begin{enumerate}
\item \emph{Axiom policies:} Given an adopting map of proof calculi $(\Phi, \alpha, \gamma) : \instP \to \instK$ on $\Thz$:
  (a)~given a functorial family of target witnesses $d = (d_\chi)_{\chi \in \alpha(\Gamma)}$ with $\pi'(d_\chi) = \chi$ in $T'$, the substitution $\gamma^d_T(p) := \gamma_T(p)[t_{d_\chi}/\instax{\chi}]$ defines a map of proof calculi (\emph{Discharge});
  (b)~over duplicate-free ordered axiom lists $\vec\varphi$ in $\instThl$, setting $\Phi_{\mathrm{rel}}(\Sigma, \vec\varphi) := (\Phi^\diamond\Sigma, \emptyset)$, $\alpha_{\mathrm{rel},T}(\psi) := \alpha(\varphi_1) \to \cdots \to \alpha(\varphi_k) \to \alpha(\psi)$, and $\rho_T(t, \psi) := (\tttlam{h_1}{\alpha(\varphi_1)}\allowbreak\cdots\allowbreak\tttlam{h_k}{\alpha(\varphi_k)}\allowbreak \gamma_T(t)[h_i/\instax{\alpha(\varphi_i)}],\,\allowbreak \alpha_{\mathrm{rel},T}(\psi))$ defines a map of proof calculi into the axiom-free fragment of $\instK$ (\emph{Relativize}).
\item \emph{Provability and model levels:} Forgetting witness terms via $\instP \mapsto \instJTh(\mathit{ent}(\instP))$ and $(\Phi, \alpha, \gamma) \mapsto \langle\Phi, \alpha\rangle$ defines a functor $\instU : \instPC \to \piIns^{\instco}$ into the category of $\pi$-institutions and comorphisms~\cite{bittencourt-rios2022,fiadeiro-sernadas1988}. Composing $\instU$ with the Mariano--Pinto functor~\cite{mariano-pinto2017} $\instG : \piIns^{\instco} \to \Ins^{\instco}$---which defines $\Mod^{\instJ}(\Sigma) := \{C_\Sigma(\Gamma) \mid \Gamma \subseteq \Sen(\Sigma)\}$, $\Mod^{\instJ}(f) := \Sen(f)^{-1}$, and $\beta_\Sigma := \alpha_\Sigma^{-1}$---embeds $\piIns^{\instco}$ as a full coreflective subcategory of $\Ins^{\instco}$ ($\instG \dashv \instF$, $\instF \circ \instG = \mathrm{Id}$) in which every comorphism automatically satisfies the Goguen--Burstall satisfaction condition $m' \models'_{\Phi\Sigma} \alpha_\Sigma(\varphi) \iff \alpha_\Sigma^{-1}(m') \models_\Sigma \varphi$.
\end{enumerate}
\end{theorem}

\subsection{Internalized Justification Logic}
\label{sec:inst-jl}

While a kernel proof $t : P$ is \emph{factive} (establishing $P$), an empirical record of \emph{grounds} must meet two opposing demands: (i)~\emph{non-factive coexistence}---recording conflicting experimental reports ($\Phi$ from Lab~A and $\neg\Phi$ from Lab~B) must not prove $\tttFalse$ as axioms in $\Prop$ would; and (ii)~\emph{computable auditability}---because $\Prop$ forbids large elimination on non-subsingletons, functions computing $\tttBool : \Setu$ (such as $\jlsw$) cannot inspect $\Prop$ proofs directly. Eigenius therefore internalizes a Curry-style adaptation of Artemov's Justification Logic~\cite{artemov2001,artemov-fitting2019} stratified across $\Setu$ and $\Prop$.

\paragraph{Term algebra, syntactic codes, and hyperintensionality.}
Let $\jlName : \Setu$ be resource IRIs and $\jlCode : \Setu$ be binary constructor trees encoding raw closed kernel terms, with quotation $\jlq{a} : \jlCode$. The unindexed (Curry-style) inductive type $\JTerm : \Setu$---kept unindexed so a pooled dossier $\jsum(s, t)$ can simultaneously justify distinct propositions $P$ and $Q$---is generated by:
\[
  j ::= \jdecl(r) \mid \jobs(r) \mid \jver(r) \mid \japp(s, t) \mid \jsum(s, t) \mid \jinst(j, \jlq{a})
\]
(plus optional $\jpair(s, t)$ and $\jlchk j$). Leaves $\lambda \in \jlLeaf$ tag epistemic kinds: $\jdecl(r)$ (declared bridge), $\jobs(r)$ (pinned readout), and $\jver(r)$ (checked kernel proof).

A chain-controlled \emph{constant specification} $\jlcs : \jlLeaf \rightharpoonup \jlCode$ maps each admitted leaf $\lambda$ to the code $\jlq{P_\lambda}$ of a closed proposition ($\Sigma \vdash P_\lambda : \Prop$), subject to the invariant that every admitted $\jver(r)$ leaf has a checked kernel proof ($\Sigma \vdash t_r : P_{\jver(r)}$). Because $\JTerm$ and $\jlCode$ live in $\Setu$, justification terms and proposition codes are first-class data. The \emph{justification assertion} $\JB : \JTerm \to \jlCode \to \Prop$ (written $\JB(j, P)$ for $\JB(j, \jlq{P})$) is an internal proposition stating that $j$ justifies $P$. Indexing $\JB$ by $\jlCode$ rather than $\Prop$ is forced by \emph{hyperintensionality}: for an admitted leaf $\lambda$ with $P_\lambda = P$, $\JB(\lambda, P)$ holds whereas $\JB(\lambda, P \wedge P)$ fails because $P \not\defeq P \wedge P$, even though $P \leftrightarrow P \wedge P$ is provable; in any extension with propositional extensionality, a predicate indexed directly by $\Prop$ could not distinguish them.

\paragraph{Evidence function and decision procedure.}
Up to definitional equality $\defeq$ on weak-head normal forms, the \emph{minimal evidence function} $\Evid_{\jlcs}(j)$ computes the finite set of closed propositions justified by $j$:
\begin{align*}
  \Evid_{\jlcs}(\lambda) &= \{P_\lambda\} \text{ if } \lambda \in \mathrm{dom}(\jlcs),\text{ else }\emptyset, \\
  \Evid_{\jlcs}(\japp(s, t)) &= \{B \mid (A \jlimp B) \in \Evid_{\jlcs}(s),\ A \in \Evid_{\jlcs}(t)\}, \\
  \Evid_{\jlcs}(\jsum(s, t)) &= \Evid_{\jlcs}(s) \cup \Evid_{\jlcs}(t), \\
  \Evid_{\jlcs}(\jinst(j, \jlq{a})) &= \{B[a/x] \mid (\Pi x{:}T.\,B) \in \Evid_{\jlcs}(j),\ \Sigma \vdash a : T\}.
\end{align*}
A deductive derivation system $\jlvd \JB(j, P)$ is generated by the rules \tttrule{Leaf}, \tttrule{App}, \tttrule{SumL}, \tttrule{SumR}, \tttrule{Inst}, and \tttrule{Conv} ($P \defeq P'$).

\begin{theorem}[Decision procedure, non-factivity, and conservativity]\label{thm:jl-main}
\leavevmode
\begin{enumerate}
\item \emph{Soundness, completeness, and complexity:} For every $j : \JTerm$ with $n$ tree nodes and closed $P : \Prop$, $|\Evid_{\jlcs}(j)| \le |\mathrm{leaves}(j)| \le n$ (\code{card\_evid\_le}), and $\jlvd \JB(j, P) \iff P \in \Evid_{\jlcs}(j)$ (Mkrtychev basic semantics~\cite{mkrtychev1997}), decidable in $O(n^2)$ kernel conversion/checking calls.
\item \emph{Non-factivity:} For any admitted $\jdecl$ or $\jobs$ leaf $\lambda$ whose proposition $P_0$ does not mention $\JB$ and is not provable in the base kernel $K$, the implication $\JB(\lambda, P_0) \jlimp P_0$ is unprovable in $K_{\JB}$.
\item \emph{Syntactic conservativity:} The extension $K_{\JB}$ of the kernel type theory $K$ by $\JB$ and its closed decision axioms ($\JB(j_0, c_0)$ when $P \in \Evid_{\jlcs}(j_0)$ and $\neg\JB(j_0, c_0)$ otherwise) is a conservative extension of $K$: every proof of a $\JB$-free proposition $\varphi$ in $K_{\JB}$ translates via a signature morphism into a proof of $\varphi$ in $K$.
\end{enumerate}
\end{theorem}
\begin{proof}[Proof sketch]
(1)~Each $\jinst$ maps every $F \in \Evid_{\jlcs}(j_0)$ to at most one instance while $\jsum$ unions disjoint subtrees, giving $|\Evid_{\jlcs}(j)| \le |\mathrm{leaves}(j)|$; $\Pi$-injectivity and transitivity of $\defeq$ match $\Evid_{\jlcs}$ to $\jlvd$, and $\japp$ nodes compare $\le |\mathrm{leaves}(s)| \cdot |\mathrm{leaves}(t)|$ pairs, summing over LCA leaf pairs to $O(n^2)$. (3)~Given a derivation $\mathcal{D}$ in $K_{\JB}$ invoking positive decision rules at closed $(t_1, c_1), \ldots, (t_m, c_m)$, define in $K$ the finite disquotational predicate~\cite{halbach2011} $J(j, c) := \mathsf{T}(\mathsf{eq}\,\langle j, c\rangle\,\langle t_1, c_1\rangle \mathbin{\|} \cdots \mathbin{\|} \mathsf{eq}\,\langle j, c\rangle\,\langle t_m, c_m\rangle)$ via code equality $\mathsf{eq} : \jlCode \to \jlCode \to \tttBool$. Substituting $J$ for $\JB$ discharges positive axioms to $\tttlam{P}{\Prop}\tttlam{p}{P} p$ and (by confluence and determinism of $\Evid_{\jlcs}$) negative axioms to $\tttlam{h}{\tttFalse} h$ (\code{Justification/\allowbreak Conservativity.lean}), yielding~(2).
\end{proof}

\paragraph{Proposition-relative support, provenance semirings, and warrant grades.}
To audit \emph{which leaves} a justification of $P$ by $j$ depends on, we define the \emph{proposition-relative support} $\supp(j, P) \in \jlWhy(\jlLeaf)$ as a finite set of finite leaf sets (\emph{alternatives}):
\begin{align*}
  \supp(\lambda, P) &= \{\{\lambda\}\} \text{ if } \lambda \in \mathrm{dom}(\jlcs),\ P_\lambda \defeq P\text{ (else }\emptyset\text{)}, \\
  \supp(\japp(s, t), P) &= \textstyle\bigcup_{A} \bigl\{a \cup b \;\big|\; a \in \supp(s, A \jlimp P),\\[-2pt]
  &\hspace{5.5em} b \in \supp(t, A)\bigr\}, \\
  \supp(\jsum(s, t), P) &= \supp(s, P) \cup \supp(t, P), \\
  \supp(\jinst(j, \jlq{a}), P) &= \textstyle\bigcup \bigl\{\supp(j, \Pi x{:}T.\,B) \;\big|\; (\Pi x{:}T.\,B) \in \Evid_{\jlcs}(j),\\[-2pt]
  &\hspace{3.2em} \Sigma \vdash a : T,\ B[a/x] \defeq P\bigr\}.
\end{align*}
Evaluating support relative to $P$ automatically filters out \emph{phantom summands}: if $P \notin \Evid_{\jlcs}(t)$, then $\supp(\jsum(s, t), P) = \supp(s, P)$, preventing an ungrounded summand $t$ from creating a vacuous alternative during counterfactual retraction queries.

\begin{theorem}[Provenance semiring homomorphism, polynomial evaluation, and factivity of verified alternatives]\label{thm:jl-semiring}
\leavevmode
\begin{enumerate}
\item \emph{Semiring factorization:} $\supp(j, P)$ is the image of the derivation polynomial $\jlprov(j, P) = \sum_{D \in \mathrm{Der}(j,P)} m(D) \in \mathbb{N}[\jlLeaf]$ under the canonical semiring homomorphism $h : \mathbb{N}[\jlLeaf] \allowbreak\to\allowbreak \jlWhy(\jlLeaf)$~\cite{green-karvounarakis-tannen2007}. For any bounded distributive lattice $(L,\allowbreak \vee,\allowbreak \wedge,\allowbreak \bot,\allowbreak \top)$ and leaf valuation $v : \jlLeaf \to L$, the map $h_v(S) :=\allowbreak \bigvee_{a \in S} \bigwedge_{\lambda \in a} v(\lambda)$ is a semiring homomorphism $\jlWhy(\jlLeaf) \to L$ that factors through the PosBool antichain quotient $\mathsf{Min} : \jlWhy(\jlLeaf) \to \jlPos(\jlLeaf)$.
\item \emph{Decidable warrant queries in $O(n^3)$:} Although $|\mathsf{Min}(\supp(j, P))|$ can be $2^{\Omega(n)}$ for a term of size $n$, every homomorphic query $g(\supp(j, P))$---such as (a)~all-verified status $\jlfv(j, P)$, (b)~counterfactual survival $\jlsw(j, P, r)$ ($\exists a \in \supp(j, P).\allowbreak\,r \notin \mathrm{res}(a)$), or (c)~the warrant grade $h_{\mathrm{kind}}(\supp(j, P))$ over $\bot < \jdecl < \jobs < \jver$---is computable in $O(n^3)$ semiring operations by dynamic programming over $(u, Q \in \Evid_{\jlcs}(u))$ without materializing $\supp(j, P)$.
\item \emph{Scalar grades lose counterfactuals, while $\jlfv$ recovers factivity:} Two terms $j_1 = \jobs(r_1)$ and $j_2 = \jobs(r_2)$ ($r_1 \neq r_2$) have identical grades under every leaf-kind valuation yet satisfy $\jlsw(j_1, P, r_1) = \mathsf{false} \allowbreak\neq\allowbreak \mathsf{true} = \jlsw(j_2, P, r_1)$. Conversely, if $\jlfv(j, P)$ holds (some $a \in \supp(j, P)$ consists solely of $\jver$ leaves), then a closed kernel proof $\Sigma \vdash t : P$ can be reconstructed recursively from $j$, $a$, and the leaf proofs $t_r$.
\end{enumerate}
\end{theorem}

\subsection{Typed Knowledge-Graph Calculus}
\label{sec:inst-kg}

Standard proof-assistant records (such as Lean structures or \emph{Unit} telescopes) compile to closed positional $\Sigma$-telescopes, which fail on open-world scientific graphs in three ways: (i)~\emph{order sensitivity} on unordered RDF/JSON-LD property sets, (ii)~\emph{closed-world fragility} and diamond-inheritance coherence obligations when curating new fields or multiple supertypes, and (iii)~\emph{mutual reference cycles} (e.g., interacting genes $\mathsf{WRN} \leftrightarrow \mathsf{FEN1}$), which inductive least fixed points reject ($\mu\kgPhi = \emptyset$). Eigenius resolves~(i)--(iii) by embedding Cooper's Type Theory with Records (TTR)~\cite{cooper2023perception} with canonicalization, coercion-free subsumption, and coinductive membership.

\paragraph{Canonical record telescopes and width/depth subsumption.}
A \emph{field set} $\Phi = \{(\ell_1, A_1), \ldots, (\ell_n, A_n)\}$ pairs distinct labels $\ell_i \in \mathcal{L}$ (under lexicographic order $<$) with types $A_i$ where labels may occur free, inducing $\ell_j \prec \ell_i \iff \ell_j \in \mathsf{FV}_{\mathcal{L}}(A_i)$. When $\prec$ is acyclic, Kahn's sort with $<$-tie-breaking yields a unique \emph{canonical telescope} $\kgcanon(\Phi) = (\ell_1 : A_1, \ldots, \ell_n : A_n)$: unordered records $\rtype{\ell_1 : A_1, \ldots, \ell_n : A_n}$ with the same field set compile to \emph{syntactically identical} telescopes ($\defeq$ by reflexivity).

Figure~\ref{fig:kg-rules} gives the rules for record types (of universe level $w \ge 1$, inhabiting $\Srt\,(w+1)$), width-inclusive record introduction (\tttrule{Rec-I}, where a record literal $\kgrcd{k = a_k}_{k \in K}$ inhabits every record type $R$ with $\kglabels(R) \subseteq K$), projection (\tttrule{Rec-E}), computation (\tttrule{Rec-$\beta$}), singleton types $\sing{a}$ (for manifest fields $\ell = a : T$), binary joins $T_1 \ttrjoin T_2$, and refinements $R \ttrmeet \bigwedge S$.

\begin{figure}[t]
\footnotesize
\begin{mathpar}
\inferrule*[right=\tttrule{Rec-F}]
  {\ctx, \ell_1{:}A_1, \ldots, \ell_{i-1}{:}A_{i-1} \vdash A_i : \Srt\,u_i\ (1 \le i \le n) \\ u_i \le w}
  {\ctx \vdash \rtype{\ell_1 : A_1, \ldots, \ell_n : A_n} : \Srt\,(w+1)}
\\
\inferrule*[right=\tttrule{Rec-I}]
  {\kglabels(R) \subseteq K \\ \ctx \vdash a_\ell : A_\ell[a_{\ell'}/\ell']\ (\ell \in \kglabels(R)) \\
   \ctx \vdash a_k : B_k \;\; \ctx \vdash B_k : \Srt\,w_k \;\; w_k \le w\ (k \in K \setminus \kglabels(R))}
  {\ctx \vdash \kgrcd{k = a_k}_{k \in K} : R}
\\
\inferrule*[right=\tttrule{Rec-E}]
  {\ctx \vdash t : R \\ \ell \in \kglabels(R)}
  {\ctx \vdash t.\ell : A_\ell[t.\ell'/\ell']_{\ell' \prec^+ \ell}}
\and
\inferrule*[right=\tttrule{Rec-$\beta$}]
  {k \in K}
  {\kgrcd{k = a_k}_{k \in K}.k \tttstep a_k}
\\
\inferrule*[right=\tttrule{Sing-I}]
  {\ctx \vdash a : T}
  {\ctx \vdash a : \sing{a}}
\and
\inferrule*[right=\tttrule{Sing-Sub}]
  {\ctx \vdash a : T \\ \ctx \vdash T \subty U}
  {\ctx \vdash \sing{a} \subty U}
\\
\inferrule*[right=\tttrule{Join-I$_i$}]
  {\ctx \vdash T_1, T_2 : \Srt\,(u+1)}
  {\ctx \vdash T_i \subty T_1 \ttrjoin T_2}
\and
\inferrule*[right=\tttrule{Join-L}]
  {\ctx \vdash T_1 \subty U \\ \ctx \vdash T_2 \subty U}
  {\ctx \vdash T_1 \ttrjoin T_2 \subty U}
\end{mathpar}
\vspace{-8pt}
\caption{Typing and subsumption rules for canonical records, singletons, and joins. In $\mathcal{K}_1$, \tttrule{Sing-Sub} and \tttrule{Join-L} restrict $U$ to graph types (headed by a class-system constant), keeping singletons and joins disjoint from $\Srt, \Pi, \Sigma$; elimination is propositional (\tttrule{Sing-E}: $\Pi x{:}\sing{a}.\,\Id\,T\,x\,a$, and \tttrule{Join-E} into $\Prop$).}
\Description{Typing and subsumption rules for canonical record types, singletons, and joins in the knowledge-graph calculus.}
\label{fig:kg-rules}
\vspace{-6pt}
\end{figure}

\paragraph{Ontology classes, coherence, and three-level membership.}
In a knowledge graph $\mathcal{G}$, each ontology class $C \in \mathcal{C}_{\mathcal{G}}$ is an intensional constant $C : \Setu$ ($C \subty \Entity$) with parent edges $\sqsubclass$ (reflexive--transitive closure $\kganc$) and own required properties $\kgReqz(C)$. Its \emph{inherited requirements} $\kgReq(C) := \bigcup \{\kgReqz(D) \mid C \kganc D\}$ define the associated level-$1$ record type $\RC{C} := \rtype{p : \kgty{p} \mid p \in \kgReq(C)}$.

\begin{theorem}[Coherence, identity-coercion soundness, and three-level membership]\label{thm:kg-metatheory}
\leavevmode
\begin{enumerate}
\item \emph{Identity-coercion soundness of \tttrule{Rec-Sub}:} If $\ctx \vdash R \subty R'$ by \tttrule{Rec-Sub} (combining width $\kglabels(R') \subseteq \kglabels(R)$, depth $\ctx, \kgtel(R) \vdash A_\ell \subty A'_\ell$, and level bound $w \le w'$), then every record literal inhabiting $R$ by \tttrule{Rec-I} inhabits $R'$ with the \emph{exact same term}, and every projection $t.\ell$ ($\ell \in \kglabels(R')$) at $R$ has a subtype of its type at $R'$.
\item \emph{Automatic class coherence:} Whenever $C \kganc D$, $\vdash \RC{C} \subty \RC{D}$ holds automatically by width subtyping (\tttrule{Width}); if $C \kgeqv D$ ($C \kganc D$ and $D \kganc C$), then $\RC{C}$ and $\RC{D}$ are syntactically identical telescopes while $C$ and $D$ remain distinct intensional constants.
\item \emph{Three-level membership:} Define the monotone operator $\kgPhi$ on subsets $M \subseteq \mathcal{R}_{\mathcal{G}} \times \mathcal{C}_{\mathcal{G}}$ by $(r, C) \in \kgPhi(M)$ iff (a)~$\exists D \in \mathcal{S}_r.\,D \kganc C$ (\emph{declared membership}) and (b)~for every $p \in \kgReq(C)$, $r$'s record $\rho_r$ has a value at $p$ satisfying $:_{M} \kgty{p}$, where an IRI reference $v$ checks at class range $D$ iff $(v, D) \in M$, lifted positively over lists, joins, singletons, and records (\emph{structural membership}). Its greatest fixed point $\nu\kgPhi$ (coinductive graph membership, admitting IRI cycles) stabilizes in at most $|\mathcal{R}_{\mathcal{G}}| \cdot |\mathcal{C}_{\mathcal{G}}|$ descending steps (\code{gfp\_eq\_iterate\_of\_finite}) and is decidable in $O(|\mathcal{R}_{\mathcal{G}}| \cdot |\mathcal{C}_{\mathcal{G}}| \cdot |\mathcal{P}_{\mathcal{G}}|)$ time by worklist counter propagation. Moreover, defining $\isa(x, C) := \exists y{:}C.\,\Id(\Entity, x, y)$ and $\subclassof(C, D) := \Pi x{:}C.\,\isa(x, D)$, every $(r, C) \in \nu\kgPhi$ yields $\vdash \tttpair{r}{\refl\,r} : \isa(r, C)$, whereas a theory may prove $\isa(r, C)$ (e.g., via $\Id(\Entity, r, s)$) without $r : C$ holding judgmentally.
\end{enumerate}
\end{theorem}

\section{Set-Theoretic Semantics \& Consistency}
\label{sec:semantics}

To establish logical consistency of the Eigenius kernel and its knowledge-graph extension, we construct a direct set-theoretic model of $\mathcal{K}_0$ and $\mathcal{K}_1$ and lift it declaratively via Theorem~\ref{thm:bidir-metatheory}(3).

\subsection{Direct Set-Theoretic Model}
\label{sec:semantics-model}

\paragraph{Overcoming the untyped-conversion / impredicativity dilemma.}
Constructing a set-theoretic model for a calculus combining \emph{untyped} $\beta$-conversion, \emph{impredicative} $\Prop$, and \emph{cumulative universes} confronts a classic obstacle~\cite{barras2010,lee-werner2011,miquel-werner2003,werner2008}. Under Aczel's trace encoding~\cite{aczel1999relating}, $\mathrm{lam}(f) = \bigcup_{(x,y) \in f} \{x\} \times y$, the empty function on $\emptyset$ denotes $\emptyset$, just like every constant-$\emptyset$ function; thus $(\tttlam{x}{\tttFalse} x)\,a$ denotes $\emptyset$ even when $\llbracket a \rrbracket = \{\emptyset\}$, violating untyped $\beta$-soundness outside the domain. Models of typed conversion~\cite{carneiro2019,lee-werner2011} avoid this by checking the argument of $\beta$ against the domain, but applying them to an untyped-conversion kernel requires equating untyped joinability with typed definitional equality---open under cumulative universes~\cite[Section~5]{lee-werner2011}. Conversely, if every function records its domain, $\mathrm{lam}(A, f) = \{(x, f(x)) \mid x \in A\}$, then under impredicativity a proof of $\Pi P{:}\Prop.\,P \to P$ would record $\llbracket\Prop\rrbracket$ while belonging to an element of $\llbracket\Prop\rrbracket$, violating Foundation in ZFC.

We resolve both obstacles simultaneously by adapting the proof-irrelevant stratified construction of \citet{miquel-werner2003} to bidirectional derivations over Grothendieck universes $V_0 \in V_1 \in \cdots$ (associated with an increasing sequence of inaccessible cardinals $\kappa_0 < \kappa_1 < \cdots$, so $V_n := V_{\kappa_n}$ is transitive and $V_0 \subseteq V_1 \subseteq \cdots$):
\begin{enumerate}
\item \textbf{Sort agreement from bidirectional derivations.} In $\mathcal{K}_0$ and $\mathcal{K}_1$, a term $t$ in context $\ctx$ (at a level valuation $v$) is classified as a \emph{proof} if some term with the same erasure $|t|$ (up to $\sim_{\mathrm{lvl}}$) checks, in a context joinable with $\ctx$, against a type formed at a universe whose level evaluates to $0$; similarly, $t$ is a \emph{proposition} if some such term infers a universe of level $0$. Because $\Prop$ is isolated from data universes ($\lzero \cumle n \iff n \tttsim \lzero$), we prove \emph{Sort Agreement} (\code{TypeTheory/KSorts.lean}): all derivations for a typable term agree on whether it is a proof or a proposition, both properties depend on $t$ only through $|t|$ up to $\sim_{\mathrm{lvl}}$ and are preserved by annotated reduction $\tttstep_a$, and no type is a proof.
\item \textbf{Stratified interpretation $\llbracket t \rrbracket_\ctx\rho$.} Semantic types are plain sets ($\mathrm{El}(A) = A$) with no tags:
  \begin{itemize}
  \item $\llbracket\Prop\rrbracket = \mathcal{P}(\{\emptyset\}) = \{\emptyset, \{\emptyset\}\}$, $\llbracket\Srt\,(\lsucc\,\ell)\rrbracket = V_{\llbracket\ell\rrbracket_v}$, and every \emph{proof} term $t$ (as well as every $\refl\,a$) denotes $\emptyset$;
  \item A data abstraction $\tttlam{x}{A} b$ (which is not a proof) denotes the explicit domain-recording graph $\mathrm{lam}(\llbracket A \rrbracket_\ctx\rho,\, v \mapsto \llbracket b \rrbracket_{\ctx, x{:}A}\rho[x \mapsto v])$, with $\mathrm{app}(\mathrm{lam}(A, f), x) = f(x)$ for $x \in A$ (undefined for $x \notin A$);
  \item Propositional products/sums denote $\widehat\Pi_{\mathrm{p}}(A, F) = \{y \in \{\emptyset\} \mid \forall x \in A.\,\emptyset \in F(x)\}$ and $\widehat\Sigma_{\mathrm{p}}(A, F) = \{y \in \{\emptyset\} \mid \exists x \in A.\,\emptyset \in F(x)\}$ in $\mathcal{P}(\{\emptyset\})$, while data products/sums denote $\widehat\Pi_{\mathrm{d}}(A, F) = \{\mathrm{lam}(A, f) \mid \forall x \in A.\,f(x) \in F(x)\}$ and $\widehat\Sigma_{\mathrm{d}}(A, F) = \{(a, b) \mid a \in A,\ b \in F(a)\}$ in $V_{\max(\ell_1, \ell_2)}$;
  \item $\llbracket\Id\,A\,a\,b\rrbracket_\ctx\rho = \{\emptyset\}$ if $\llbracket a \rrbracket_\ctx\rho = \llbracket b \rrbracket_\ctx\rho$ (else $\emptyset$); a non-proof $\Jelim\,a\,C\,d\,b\,p$ denotes $\llbracket d \rrbracket_\ctx\rho$; and $\llbracket\tttann{u}{U}\rrbracket_\ctx\rho = \llbracket u \rrbracket_\ctx\rho$;
  \item An admissible inductive family $D$ denotes the Knaster--Tarski least fixed point \code{indFam} of its monotone constructor operator (in $V_\ell$ for data, stabilizing below $\kappa_\ell$ by strict positivity, and in $\mathcal{P}(\{\emptyset\})$ for $\Prop$). In \code{SIndRec.lean}, $\rec_D$ is interpreted via the inductive relation \code{recGraph} of its semantic $\iota$-clauses: \emph{functionality} (\code{recGraph\_functional}) holds by induction on \code{recGraph}---even for recursive subsingletons like $\tttAcc$ where every proof denotes $\emptyset$, because proof fields equal $\emptyset$, index-forced data fields equal $\bar\imath$, and recursive hypothesis codes are uniquely determined on predecessors---while \emph{totality} on $\llbracket D\,\bar p\,\bar\imath\rrbracket$ (\code{recFun\_spec}) holds by induction on the least fixed point \code{indFam} (\code{SAcc.lean}; non-well-founded open assumptions $p_0 : \tttAcc\,A\,R\,a_0$ simply have $\llbracket\ctx\rrbracket = \emptyset$).
  \end{itemize}
\end{enumerate}

Because data $\Pi$-types have $\ell_2 > 0$, their universe level is $\limax\,\ell_1\,\ell_2 = \max(\ell_1, \ell_2)$, so the domain-recording graphs $\mathrm{lam}(A, f)$ are ordinary sets in $V_{\max(\ell_1, \ell_2)}$ and never encounter impredicativity.

\begin{definition}[Single environment obligation]\label{def:env-obligation}
A well-formed global environment $\tttEnv$ \emph{meets its semantic obligation} if for every axiom $\mathsf{axiom}\ \tttinst{c}{\bar u} : A$ in $\tttEnv$ and every level valuation of $\bar u$, $\llbracket A \rrbracket$ is defined and $\llbracket A \rrbracket \neq \emptyset$ in the model of the preceding prefix. (Definitions and admissible inductive declarations carry no obligation.)
\end{definition}

In particular, the proof-irrelevance axiom $\irrel$ meets Definition~\ref{def:env-obligation} in \emph{every} model: for any $P \in \mathcal{P}(\{\emptyset\})$ and $p, q \in P$, we have $p = \emptyset = q$, so $\llbracket\Id\,P\,p\,q\rrbracket = \{\emptyset\}$, and $\emptyset$ inhabits the propositional product $\llbracket\Pi P{:}\Prop.\,\Pi p{:}P.\,\Pi q{:}P.\,\Id\,P\,p\,q\rrbracket = \{\emptyset\}$.

\begin{lemma}[Directional soundness of annotated reduction]\label{lem:strat-soundness}
Over any $\tttEnv$ meeting Definition~\ref{def:env-obligation}:
\begin{enumerate}
\item If $\ctx \vdash t \infr T$ or $\ctx \vdash t \chk T$ in $\mathcal{K}_1$, $t \tttstep_a t'$, and $\llbracket t \rrbracket_\ctx\rho$ is defined, then $\llbracket t' \rrbracket_\ctx\rho = \llbracket t \rrbracket_\ctx\rho$.
\item If $\ctx \vdash t \chk T$, $\ctx \vdash t' \chk T'$, $t \defeq_K t'$, and both $\llbracket t \rrbracket_\ctx\rho$ and $\llbracket t' \rrbracket_\ctx\rho$ are defined, then $\llbracket t \rrbracket_\ctx\rho = \llbracket t' \rrbracket_\ctx\rho$.
\end{enumerate}
\end{lemma}
\begin{proof}[Proof sketch]
Part~(1) is \emph{directional} and proved prior to typing soundness: at a data $\beta$-redex $(\tttlam{x}{A} b)\,a$, definedness of $\llbracket(\tttlam{x}{A} b)\,a\rrbracket_\ctx\rho = \mathrm{app}(\mathrm{lam}(\llbracket A \rrbracket_\ctx\rho, f), \llbracket a \rrbracket_\ctx\rho)$ already yields $\llbracket a \rrbracket_\ctx\rho \in \llbracket A \rrbracket_\ctx\rho$, breaking the circularity with typing soundness, while \tttrule{Id-$\iota$} steps $\Jelim\,a\,C\,d\,b\,(\refl\,a') \tttstep_a \tttann{d}{C\,a\,(\refl\,a)}$ preserve $\llbracket d \rrbracket_\ctx\rho$ directly without needing $\llbracket a \rrbracket_\ctx\rho = \llbracket b \rrbracket_\ctx\rho$. Part~(2) lifts $|t| \tttstep^* s$ and $|t'| \tttstep^* s$ to $\tttstep_a^*$ via Theorem~\ref{thm:bidir-metatheory}(2) and applies~(1).
\end{proof}

\subsection{Logical Consistency \& Lean 4 Mechanization}
\label{sec:semantics-consistency}

Combining Lemma~\ref{lem:strat-soundness}, Theorem~\ref{thm:bidir-metatheory}, and the class-system and TTR record semantics yields our master consistency result:

\begin{theorem}[Master consistency of $\mathcal{K}_0$, $\mathcal{K}_1$, the declarative kernel, and the knowledge-graph calculus]\label{thm:master-consistency}
In ZFC with an increasing sequence of inaccessible cardinals $\kappa_0 < \kappa_1 < \cdots$, let $\tttEnv$ be a well-formed environment (optionally extended with $\irrel$, the justification predicate $\JB$, and the signature $\Sigma_{\mathcal{G}}$ of a well-formed knowledge graph $\mathcal{G}$) that meets Definition~\ref{def:env-obligation}.
\begin{enumerate}
\item \emph{Semantic typing and coercion-free subsumption:} For every derivation $\ctx \vdash a \infr A$ or $\ctx \vdash a \chk A$ in $\mathcal{K}_0$ or $\mathcal{K}_1$ (and hence every declarative derivation $\ctx \vdash a : A$) and every valid environment $\rho$ of $\ctx$, both $\llbracket a \rrbracket_\ctx\rho$ and $\llbracket A \rrbracket_\ctx\rho$ are defined with $\llbracket a \rrbracket_\ctx\rho \in \llbracket A \rrbracket_\ctx\rho$; and every subsumption $\ctx \vdash A \subty B$ denotes set inclusion $\llbracket A \rrbracket_\ctx\rho \subseteq \llbracket B \rrbracket_\ctx\rho$.
\item \emph{Logical consistency:} In any axiom-free environment (or any environment extending it by $\irrel$, $\JB$ decision rules, and a well-formed knowledge graph $(\Sigma_{\mathcal{G}}, \emptyset)$), there is no closed term $t$ with $\cdot \vdash t : \tttFalse$.
\end{enumerate}
\end{theorem}
\begin{proof}
(1)~By induction on the $\mathcal{K}_1$ derivation (in \code{TypeTheory/\allowbreak SValidity.lean}, \code{SConsistency.lean}, and \code{SEnvKG.lean}). At \tttrule{Inf-J}, $\llbracket p \rrbracket_\ctx\rho \in \llbracket\Id\,A\,a\,b\rrbracket_\ctx\rho$ forces $\llbracket a \rrbracket_\ctx\rho = \llbracket b \rrbracket_\ctx\rho$ and $\llbracket p \rrbracket_\ctx\rho = \emptyset = \llbracket\refl\,a\rrbracket_\ctx\rho$, giving $\llbracket C\,b\,p\rrbracket_\ctx\rho = \llbracket C\,a\,(\refl\,a)\rrbracket_\ctx\rho \ni \llbracket d \rrbracket_\ctx\rho$; at \tttrule{S-$\Pi$}, $A_1 \defeq_K A_2$ gives $\llbracket A_1 \rrbracket_\ctx\rho = \llbracket A_2 \rrbracket_\ctx\rho$ by Lemma~\ref{lem:strat-soundness}(2), so $\widehat\Pi_{\mathrm{d}}(\llbracket A_1 \rrbracket, F_1) \subseteq \widehat\Pi_{\mathrm{d}}(\llbracket A_2 \rrbracket,\allowbreak F_2)$ is a genuine subset inclusion on domain-recording function graphs; \tttrule{S-Cum} holds by transitivity $V_m \subseteq V_n$; \tttrule{Class-Ent}, \tttrule{Class-Sub}, \tttrule{Sing-Sub}, \tttrule{Join-I}, and \tttrule{Join-L} hold by inclusion on resource codes, singletons, and unions in \code{SEnvKG.lean}; and \tttrule{Rec-Sub} holds on finite label maps in $V_{w-1}$ (mechanized at the TTR semantic level in \code{KnowledgeGraphs/}). Declarative derivations translate into $\mathcal{K}_1$ by Theorem~\ref{thm:bidir-metatheory}(3). (2)~The zero-constructor family $\tttFalse$ denotes $\llbracket\tttFalse\rrbracket = \emptyset$.
\end{proof}

\paragraph{Lean~4 mechanization architecture.}
The metatheory is mechanized in Lean~4 (\code{v4.35.0} over Mathlib, zero \code{sorry} axioms) in \code{EigeniusTheory} (${\sim}58{,}800$ lines across 119 files; Table~\ref{tab:lean-mechanization}). Note the separation between \emph{object language} and \emph{metalanguage}: although Lean~4's own \code{Prop} enforces \tttrule{Def-PI}, Eigenius expressions---including proofs $p_{\mathrm{CRISPR}}$ and $p_{\mathrm{RNAi}}$---are values of the inductive data type \code{Term : Type}, where \code{pCRISPR} $\neq$ \code{pRNAi} holds provably; Lean's proof-irrelevant \code{Prop} is used solely for metatheorems about \code{Term}.

\begin{table}[t]
\centering\footnotesize
\setlength{\tabcolsep}{2.2pt}
\caption{Overview of the companion Lean~4 mechanization (\code{EigeniusTheory}, 119 files, ${\sim}58{,}800$ lines, zero \code{sorry} axioms).}
\label{tab:lean-mechanization}
\begin{tabularx}{\columnwidth}{@{}l >{\raggedright\arraybackslash}p{2.26cm} >{\raggedright\arraybackslash}X@{}}
\toprule
\textbf{Sec.} & \textbf{\code{EigeniusTheory/}} & \textbf{Key Mechanized Modules \& Theorems} \\
\midrule
\S\ref{sec:kernel} & \code{TypeTheory/} & \code{Level} (\code{CumLe}), \code{Confluence} (\code{confluent}), \code{DefEq} (\code{pi\_inj}, \code{ne\_of\_head\_ne}), \code{Inductive} (\code{forcedIndex}), \code{Irrel}, \code{Inconsistency} \\
\S\ref{sec:subsumption} & \code{TypeTheory/}, \code{Checking/} & \code{Bidirectional} (\code{Inf.sound}), \code{AnnotatedReduction}, \code{Completeness} (\code{HasType.translate}), \code{KSorts} (\code{sortsMotive}), \code{Conversion} (\code{ConvRun.correct}), \code{NbE} (\code{Nbe.decides}), \code{Procedure} (\code{Run.correct}), \code{Termination} (\code{ConvRun.total}) \\
\S\ref{sec:inst-env} & \code{Institutions/} & \code{ProofCalculus}, \code{Kernel} (\code{KCalc}, \code{equiv\_iff}), \code{KernelPolicies}, \code{Models} (\code{toInstitution}) \\
\S\ref{sec:inst-jl} & \code{Justification/} & \code{Evidence} (\code{jb\_iff\_mem\_evid}), \code{Support} (\code{supp\_sum\_of\_notMem}, \code{hv\_minimals}), \code{Conservativity} (\code{HasType.conservative}) \\
\S\ref{sec:inst-kg} & \code{KnowledgeGraphs/} & \code{Canonical} (\code{canon\_congr}), \code{Graph} (\code{rsub\_RC\_of\_anc}, \code{gfp\_eq\_iterate\_of\_finite}), \code{Propositions} \\
\S\ref{sec:semantics} & \code{TypeTheory/S*} & \code{SSet} (\code{IsGrothendieck}), \code{SModel}, \code{SIndRec} (\code{recFun\_spec}), \code{SAcc} (\code{consistency\_acc}), \code{SConsistency} (\code{consistency\_of\_model}), \code{SEnvKG}, \code{SUniverses} (\code{consistency\_false\_univ}) \\
\bottomrule
\end{tabularx}
\end{table}

In \code{SSet.lean} and \code{SConsistency.lean}, $V_0 \in V_1 \in \cdots$ is formalized via \code{IsGrothendieck (V : ZFSet.\{u\})} over Mathlib's Aczel sets (\code{HasType.\allowbreak consistency\_\allowbreak of\_model}). Below $\Srt\,4$ ($\Prop$ through $\Typeu_2$), \code{SUniverses.lean} builds Grothendieck universes from Lean~4's \code{Type 0..2} (\code{univCard\_\allowbreak inaccessible}), proving consistency (\code{ChkB.\allowbreak consistency\_\allowbreak false\_univ}) \emph{unconditionally} with zero set-theoretic axioms.

\section{Related Work \& Conclusion}
\label{sec:related}

\paragraph{Proof irrelevance, Coq's $\Prop$, and $\mathsf{SProp}$.}
Coq's default $\Prop$~\cite{coq-refman} shares untyped $\beta\delta\iota$-conversion and can postulate \code{ProofIrrelevance} ($\irrel$), whose set-theoretic semantics and choice hazards were analyzed by \citet{werner2008}; however, Coq makes $\Prop \subty \Setu \subty \Typeu_i$ cumulative (breaking untagged Sort Agreement, \S\ref{sec:semantics}) and builds $\eta$ into conversion. Modal irrelevance with type-directed equality in predicative type theory was studied by \citet{abel2012irrelevance}, while \citet{gilbert2019} introduced strict $\mathsf{SProp}$ (Coq, Agda's $\Prop$) and \citet{carneiro2019} specified Lean's \tttrule{Def-PI} kernel. Because \tttrule{Def-PI} with impredicative $\Id$- or $\tttAcc$-elimination breaks normalization~\cite{abel-coquand2020,gilbert2019} (Proposition~\ref{prop:def-pi-failure}), Coq forbids recursive large $\mathsf{SProp}$ elimination and \citet{felicissimo2026acc} restrict $\tttAcc$; isolating $\Prop$ ($\Prop \not\subty \Setu$) with untyped conversion and $\irrel$ retains full subsingleton elimination, confluence, and distinct witnesses.

\paragraph{Observational Type Theory.}
Observational Type Theory (OTT)~\cite{altenkirch2007ott} and $\mathsf{CC}^{\mathrm{obs}} / \mathsf{CIC}^{\mathrm{obs}}$~\cite{pujet-tabareau2023} equip impredicative $\Prop$ with structural equality and homogeneous casts reducing on matching type formers. While giving extensionality computationally, they still enforce \tttrule{Def-PI} in conversion (erasing $p_{\mathrm{CRISPR}}$ vs.\ $p_{\mathrm{RNAi}}$), whereas $\Jelim$-transport along $\irrel\,P\,p\,q$ in Eigenius stays intentionally neutral to record the exact bridge.

\paragraph{Minimalist bidirectional checkers and verified kernels.}
Following \citet{coquand1996}, \emph{Mini-TT}~\cite{minitt} and \emph{Unit}~\cite{coquand2026unit} eschew unification, tactics, and heuristics in the TCB; \emph{Lean4Lean}~\cite{carneiro2024} verifies Lean~4's \tttrule{Def-PI} kernel. Eigenius adopts \emph{Unit}'s bidirectional discipline (\S\ref{sec:subsumption-unit}, Table~\ref{tab:unit-comparison}) while omitting $\eta$ on annotated $\lambda$, isolating impredicative $\Prop$ with $\irrel$, and mechanizing WHNF conversion, untyped NbE~\cite{aehlig-joachimski2004}, and consistency.

\paragraph{Cumulative universes, record calculi, and coercive subtyping.}
Our cumulative universes and codomain-covariant $\Pi$-subsumption build on Luo's ECC~\cite{luo1994} and Pollack's bidirectional analyses~\cite{pollack2002}. Unlike coercive subtyping for Modern Type Theories~\cite{chatzikyriakidis-luo-2020,luo2009records}---where $A \le_c B$ inserts coercions with path-coherence obligations---every judgmental subsumption rule in Eigenius (\tttrule{S-Cum}, \tttrule{S-$\Pi$}, \tttrule{Rec-Sub}, \tttrule{Class-Sub}, \tttrule{Sing-Sub}, \tttrule{Join-I}/\tttrule{L}) acts as the \emph{identity on terms} and denotes set inclusion (\S\ref{sec:semantics}), keeping contravariant domain adaptation explicit as $\eta$-expansion outside $\subty$ (Proposition~\ref{prop:eta-contra}). Our embedding of Cooper's TTR~\cite{cooper2023perception} resolves the positional sensitivity of $\Sigma$-telescopes~\cite{betarte2000} via Kahn canonicalization (Theorem~\ref{thm:kg-metatheory}).

\paragraph{Justification Logic and provenance semirings.}
Artemov's Logic of Proofs ($\mathsf{LP}$) and its relatives ($\mathsf{J}, \mathsf{J4}, \mathsf{JT}$)~\cite{artemov2001,fitting2005,mkrtychev1997} replace modal $\Box P$ with evidence assertions $t : P$, while typed $\lambda$-calculi for justification logic~\cite{artemov-bonelli2007,pouliasis-primiero2014} study Church-style evidence terms. Eigenius internalizes a Curry-style algebra $\JTerm : \Setu$ inside dependent type theory, proves that Church-style indexing splits $\mathsf{LP}$'s Sum while losing hyperintensionality, and links support $\supp(j, P)$ to provenance semirings~\cite{green-karvounarakis-tannen2007}.

\paragraph{Institutions and formal knowledge management.}
Goguen and Burstall's theory of institutions~\cite{goguen1992}, Meseguer's general logics and proof calculi~\cite{meseguer1989}, Mossakowski et al.'s Heterogeneous Tool Set (Hets)~\cite{mossakowski2007hets}, and Rabe and Kohlhase's MMT framework~\cite{rabe2013mmt} provide the categorical foundation for logic translations. Eigenius extends this lineage to verified empirical repositories by combining Meseguer's witness-level proof calculi with the Mariano--Pinto coreflection~\cite{mariano-pinto2017}, classifying cross-engine comorphisms by their axiom policies (\emph{Adopt}, \emph{Discharge}, \emph{Relativize}) and linking witness transport directly to warrant monotonicity in justification logic.

\paragraph{Conclusion.}
When autonomous agents and computational pipelines generate scientific claims at scale, verifying that a proposition is inhabited is no longer enough: while empirical observations and domain bridges remain fallible, a repository can---and must---verify the inferential glue connecting them, track across which institutions evidence was transported, and audit whether conclusions survive premise retractions. Separating orthogonal untyped conversion from witnessed propositional proof irrelevance, unifying universe and ontology subsumption, and embedding justification logic and TTR knowledge graphs over constructive institutions gives such stores a mechanically verified foundation.

\begin{acks}
The author would like to thank Allen L.~Brown, Jr.\ for formulating the mission of enforcing the scientific method in Pharos~\cite{brown2008pharos}, and for being a mentor and discussion partner over the years.
The author supplied the design of Eigenius and directed its development and associated experimentation.
Claude (Anthropic) produced the implementation code, and Claude and Gemini (Google) produced the formalization code through iterative work under that direction. This attribution concerns code production; the mathematical sources are cited in the text. The resulting text and software remain the author's responsibility.
\end{acks}

\renewcommand*{\bibliofont}{\footnotesize}
\setlength{\bibsep}{0pt}

\end{document}